\documentclass[twocolumn,
english,
aps,
superscriptaddress,
amsmath,
amssymb,
floats,
prb,
longbibliography]{revtex4-2}
\usepackage[latin9]{inputenc}
\usepackage{color}
\usepackage{mathtools}
\usepackage{amsmath}
\usepackage{graphicx}
\usepackage{multirow,booktabs}
\usepackage{array}
\usepackage{xcolor,colortbl}
\usepackage[normalem]{ulem}
\usepackage{lipsum}
\usepackage{soul}
\usepackage{dsfont}

\makeatletter

\usepackage{comment}

\usepackage[unicode=true,
 bookmarks=false,
 breaklinks=true,pdfborder={0 0 1},backref=false,colorlinks=true]
 {hyperref}
\hypersetup{pdfcreator={},pdfproducer={LaTeX with hyperref},linkcolor=blue,anchorcolor=blue,citecolor=blue,filecolor=red,menucolor=red,urlcolor=blue,pdfstartview=FitV,pdfhighlight=/I,hypertexnames=true}

\@ifundefined{textcolor}{}{%
 \definecolor{BLACK}{gray}{0}
 \definecolor{WHITE}{gray}{1}
 \definecolor{RED}{rgb}{1,0,0}
 \definecolor{GREEN}{rgb}{0,1,0}
 \definecolor{BLUE}{rgb}{0,0,1}
 \definecolor{CYAN}{cmyk}{1,0,0,0}
 \definecolor{MAGENTA}{cmyk}{0,1,0,0}
 \definecolor{YELLOW}{cmyk}{0,0,1,0}
}

\newcommand{\mbf}[1]{\mathbf{#1}}

\usepackage{epstopdf}
\usepackage{array}
\usepackage{mathrsfs}
\usepackage{dcolumn}
\usepackage{bm}

\usepackage{xcolor}

\newcolumntype{M}[1]{>{\centering\arraybackslash}m{#1}}
\newcolumntype{P}[1]{>{\centering\arraybackslash}p{#1}}

\usepackage{array}
\newcommand{\PreserveBackslash}[1]{\let\temp=\\#1\let\\=\temp}
\newcolumntype{C}[1]{>{\PreserveBackslash\centering}p{#1}}
\newcolumntype{R}[1]{>{\PreserveBackslash\raggedleft}p{#1}}
\newcolumntype{L}[1]{>{\PreserveBackslash\raggedright}p{#1}}

\graphicspath{{../Inkscape/},{../Mathematica/},{./Figures/}}

\usepackage{tikz}
\usetikzlibrary{calc,intersections}
\usetikzlibrary{shadings}
\usepgflibrary{shadings}

\usepackage{babel}

\begin{document}

\title{Microscopic origin of the nemato-elastic coupling and dynamics of hybridized collective nematic-phonon excitations}
\author{Morten H. Christensen}
\email{mchriste@nbi.ku.dk}
\affiliation{Niels Bohr Institute, University of Copenhagen, 2200 Copenhagen, Denmark}

\author{Michael Sch{\"u}tt}
\affiliation{Carl Zeiss SMT GmbH, 73447 Oberkochen, Germany}

\author{Avraham Klein}
\affiliation{Physics Department, Ariel University, Ariel 40700, Israel}

\author{Rafael M. Fernandes}
\affiliation{Department of Physics, The Grainger College of Engineering, University of Illinois Urbana-Champaign, Urbana, Illinois 61801, USA}
\affiliation{Anthony J. Leggett Institute for Condensed Matter Theory, The Grainger College of Engineering, University of Illinois Urbana-Champaign, Urbana, Illinois 61801, USA}

\date{\today}
\begin{abstract}
Electronically-driven nematic order breaks the rotational symmetry of a system, e.g., through a Pomeranchuk instability of the Fermi surface, with a concomitant distortion of the lattice. As a result, in a metal, the nematic collective mode interacts with two different sets of gapless excitations: the particle-hole excitations of the metal and the lattice fluctuations that become soft at the induced structural transition, namely, the transverse acoustic phonons. However, the \textit{dynamics} of these hybridized collective modes formed by the transverse acoustic phonons and the metallic electronic-nematic fluctuations has remained largely unexplored. Here we address this problem by developing a formalism in which the nemato-elastic coupling is obtained microscopically from the direct coupling between electrons and transverse acoustic phonons enabled by impurities present in the crystal. We then demonstrate the emergence of hybrid nemato-elastic modes that mix the characteristics of the transverse phonons and of the nematic fluctuations. Near the nematic quantum critical point in a metal, two massless modes emerge with intertwined dynamic behaviors, implying that neither mode dominates the response of the system. We systematically study the non-trivial dependence of these collective modes on the longitudinal and transverse momenta, revealing a rich landscape of underdamped and overdamped modes as the proximity to the quantum critical point and the strength of the electron-phonon coupling are changed. Since dynamics play an important role for determining superconducting instabilities, our results have implications for the study of pairing mediated by electronic nematic fluctuations. 
\end{abstract}
\maketitle

\section{Introduction}

Electronic nematicity -- a quantum phase in which electrons spontaneously break rotational but not translational symmetry -- is one of the hallmarks of correlated metals~\cite{Kivelson98,Fradkin2010Nematic,Fernandes2014}. Indeed, such phases are routinely observed in a diverse range of correlated materials including iron-based superconductors~\cite{Chu2010In-plane,Yi2011Symmetry-breaking,Chu2012Divergent,Kasahara2012Electronic,Kuo2016Ubiquitous,Coldea2021,Bohmer2022}, cuprates~\cite{Hinkov2008Electronic,Lawler2010Intra-unit-cell,Bozovic2017,Auvray2019Nematic}, quantum Hall systems~\cite{Lilly1999Evidence,Feldman2016Observation}, and kagome metals~\cite{Drucker2024Incipient}. In the unconventional superconductors, the observation of a superconducting dome which seemingly straddles the putative quantum critical point (QCP) associated with the nematic transition, particularly in iron-based superconductors~\cite{Fernandes2022iron,Worasaran2021,Hosoi2016,SilvaNeto2025}, has inspired the idea of high-temperature superconductivity mediated or enhanced by nematic fluctuations~\cite{Lederer2015Enhancement,Schattner2016Ising,Lederer2017Superconductivity,Labat2017Pairing,Klein2018,Chubukov2024}. In hypothetical electronic fluids without a lattice, the nematic transition can be described as a Pomeranchuk instability of the Fermi surface which lowers the continuous rotational symmetry of the electronic liquid~\cite{Pomeranchuk1958On,Oganesyan01,Chubukov2018Fermi-Liquid}. Here, the long-wavelength nature of the near-critical nematic fluctuations serve to enhance superconducting pairing interactions leading to a superconducting dome centered on the nematic QCP~\cite{Lederer2015Enhancement}. 

The presence of a lattice complicates matters. As a result of the coupling between electrons and phonons, the electronic nematic transition is accompanied by a concomitant structural transition which lowers the symmetry of the underlying lattice. In contrast to the Pomeranchuk instability in an electronic fluid, structural transitions are not associated with critical modes but rather with the vanishing of the velocity of specific transverse acoustic phonon modes~\cite{Cowley1976Acoustic}. This tends to ``lock'' the transition to the lattice, rendering it Ising-like in the case of a tetragonal lattice rather than a continuous symmetry-breaking transition~\cite{Hecker2024}. In fact, for many structural transitions, the velocity only vanishes along certain directions~\cite{Cowley1976Acoustic}. Since phonons are Goldstone modes of the lattice, the coupled nematic-structural transition is also not accompanied by a critical mode but, as above, by the vanishing of the velocity of a transverse acoustic phonon along certain directions. Crucially, this implies that the coupling between the nematic fluctuations and the lattice cannot be neglected~\cite{Karahasanovic16,Paul17,Carvalho2019,Carvalho2022,Fernandes2020,Hecker2022,Labat2017Pairing,Massat2022}. Indeed, as nematic fluctuations are electronic in origin, they necessarily involve a coupling of electrons to transverse phonons. As such, the impact of the transverse phonons on the nematic fluctuations depend crucially on the electronic density and, in particular, metals and insulators will exhibit distinct behaviors. In this manuscript, we will focus on metallic systems.

\begin{figure}[t]
    \centering
    \includegraphics[width=0.95\columnwidth]{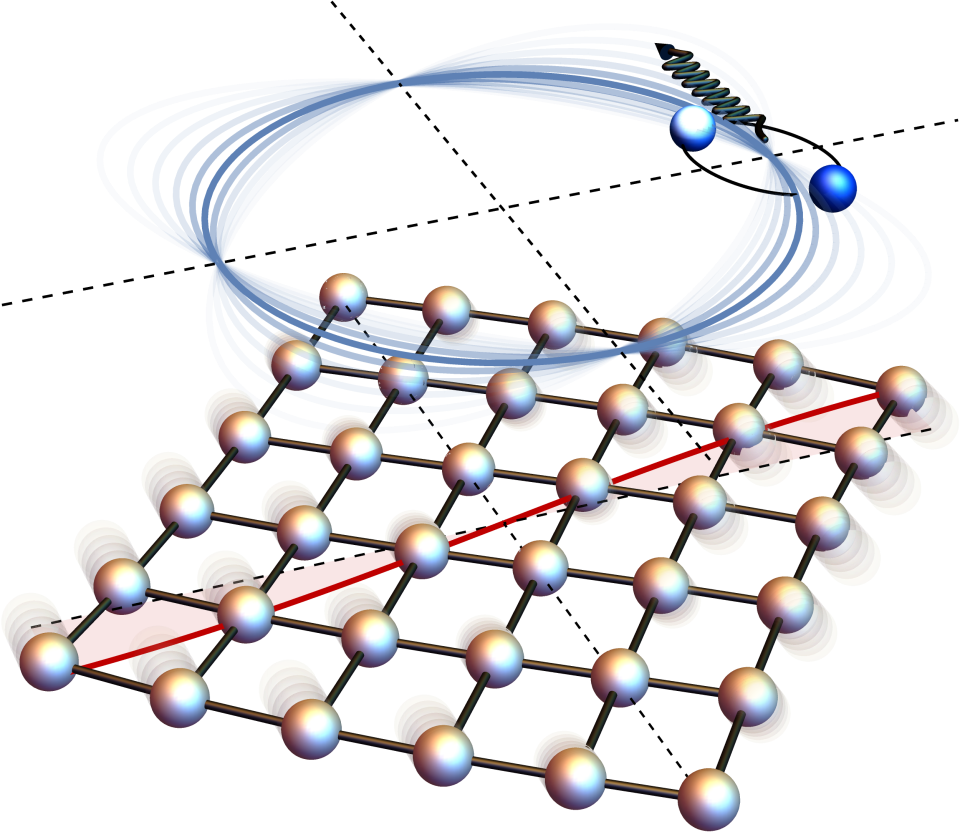}
    \caption{\label{fig:hot_and_cold_spots} \textbf{Schematic electronic-nemato-elastic system.} The nematic (i.e., quadrupolar) fluctuations of the Fermi surface allows the formation of underdamped particle-hole excitations with momentum slightly off the diagonal directions as the QCP is approached~\cite{Oganesyan01}. Further away from the diagonal directions, fluctuations become overdamped and rapidly decay into the particle-hole continuum. The transverse phonons soften precisely along the diagonal directions and the directional dependence of the nematic fluctuations implies that the two degrees of freedom strongly hybridize. As a result of this hybridization, the nematic mode becomes damped even at the QCP, while a new mixed massless coherent mode emerges.}
\end{figure}

The ``traditional'' pathway to analyzing the behavior of itinerant electrons near a symmetry breaking transition, such as the Ising-nematic one considered in this work, is to artificially break up the electron dynamics into that of the (fermionic) Fermi-liquid quasiparticles, and that of an emergent (bosonic) collective excitation. The collective excitation is then represented by a fermionic bilinear, e.g. $\phi(\mathbf{q},t) = \sum_{\mathbf{k}}\bar{\psi}_{\mathbf{k}+\mathbf{q}/2}(t)\cos(2\theta_{\mbf{k}})\psi_{\mathbf{k}-\mathbf{q}/2}(t)$ for the Ising-nematic operator with $d_{x^2-y^2}$ symmetry, where $\theta_{\mathbf{k}}$ is the angle between $\mathbf{k}$ and the major lattice axis. The emergent physics is captured in the low-energy theory by the coupling between the quasiparticles and the collective excitation, and the transition occurs when the gap for the emergent nematic field, $r_\phi$, softens to zero. In this picture, the coupling between the nematic mode and the electrons naturally vanishes along the diagonals, as evinced by the vanishing of the $\cos(2\theta_{\mbf{k}})$ term above, creating ``cold spots'' that are unaffected by the transition. However, such a picture neglects that the lattice must also break its symmetry at the phase transition, and hence the electronic nematic mode $\phi$ necessarily couples to an elastic mode -- in our case, an acoustic mode associated with the lattice displacement $\mathbf{u}$. Crucially, this acoustic mode softens only along the diagonals, and hence its contribution to the thermodynamics of the electronic mode is maximal along the diagonals. To account for this, one may integrate out the phonon degrees of freedom to obtain the renormalization of the nematic mode. What is found is that the nematic gap is modified to $r_{\ast} = r_{\phi} - \lambda^2/v_s^2$, where $\lambda$ is the electron-phonon coupling and $v_s$ is the phonon velocity, but \emph{only} along the diagonals~\cite{Paul17,Carvalho2019}. Thus, only electrons located near the ``cold spots'', corresponding to the intersection between the Fermi surface and the diagonals, can exchange low-energy soft nematic fluctuations.

The picture emerging from this argument is that when considering the thermodynamics of the electronic nematic system, the Fermi-surface points at the diagonals are not the cold but the ``hot'' spots. Since precisely there the nematic-electron coupling is small, the diagonals are ``cold'' for electrons and ``hot'' for the nematic fluctuations, and e.g. superconductivity is only weakly affected by the nematic mode \cite{Labat2017Pairing}. Unfortunately, such a picture is also incomplete near a quantum phase transition, which is governed not just by the low-energy, quasistatic behavior of the system but also by its dynamics. Hence, a full understanding of the interplay between nematicity and electronic fluctuations requires obtaining the \emph{dynamical} correlation function of $\phi$ in both momentum and frequency space. Importantly, because of the nematic-phonon coupling, the dynamical collective mode is actually a hybrid electronic-lattice soft excitation, implying that both electronic nematicity and phonons must be treated on an equal footing. 

In this work, we perform an analysis of the coupled electron/electronic nematic/phonon system shown in Fig.~\ref{fig:hot_and_cold_spots} and study its emergent collective modes. The dynamics of the electronic nematic order parameter is impacted by two different types of gapless excitations: the particle-hole continuum of the underlying electronic liquid and the transverse acoustic (TA) phonons of the underlying lattice. To capture the full dynamics of the system, we treat these three degrees of freedom on an equal footing, rather than integrating out the gapless modes and focusing on a renormalized nematic susceptibility. To accomplish this, we must include the microscopic coupling between the electrons and the TA phonons. This is an important difference with respect to previous approaches, in which the nemato-elastic coupling was included phenomenologically~\cite{Karahasanovic16,Paul17,Carvalho2019}. As with any transverse phonon mode, the electron-phonon matrix element vanishes in the usual approximation in which the coupling is via the gradient of the lattice potential. Instead, here we consider the electron-phonon matrix element modified by the presence of impurities. Since the latter are also displaced by the phonons~\cite{Tsuneto1961Ultrasound,Schmid1973Electron-phonon,Kamenev2017}, this results in a non-zero coupling between electrons and TA modes.  

We find that the quasistatic picture above is qualitatively modified when considering the full dynamics of the coupled electronic-nematic-lattice degrees of freedom. Our main result is that the usual soft nematic collective mode illustrated in Fig.~\ref{fig:hot_and_cold_spots} -- which in the absence of a lattice becomes underdamped as the QCP is approached \cite{Oganesyan01} -- remains damped even at the QCP once the lattice is included. Importantly, this mode remains the sharpest along the diagonal directions of the system, which is where the nematic form factor vanishes and the phonon velocity softens. Moreover, a new hybrid collective mode emerges which becomes progressively more coherent as the QCP is approached. This mode is a mixture of the original gapless nematic and phononic modes and, as a consequence of Adler's theorem, it remains massless regardless of the distance to the QCP. This behavior is consistent with recent observations of a lack of divergent effective masses near the nematic critical point in FeSe$_{1-x}$S$_x$~\cite{Reiss2020Quenched,Coldea2021Electronic}.

This manuscript is organized as follows: In Sec.~\ref{sec:adler} we briefly review the implications of Adler's theorem for the case of a phenomenological nemato-elastic system. This is followed by a summary of the dynamics of nematic fluctuations in the absence of phonons in Sec.~\ref{sec:nematic_fluctuations} and a review of the phonon dynamics near a structural phase transition in Sec.~\ref{sec:transverse_ph}. We present the coupled electronic-nematic-phononic system in Sec.~\ref{sec:coupled_system} and conclude in Sec.~\ref{sec:conclusions}. Further technical details are presented in three Appendices.

\section{Collective modes of the nemato-elastic system}\label{sec:adler}

\begin{figure}
\includegraphics[width=\columnwidth]{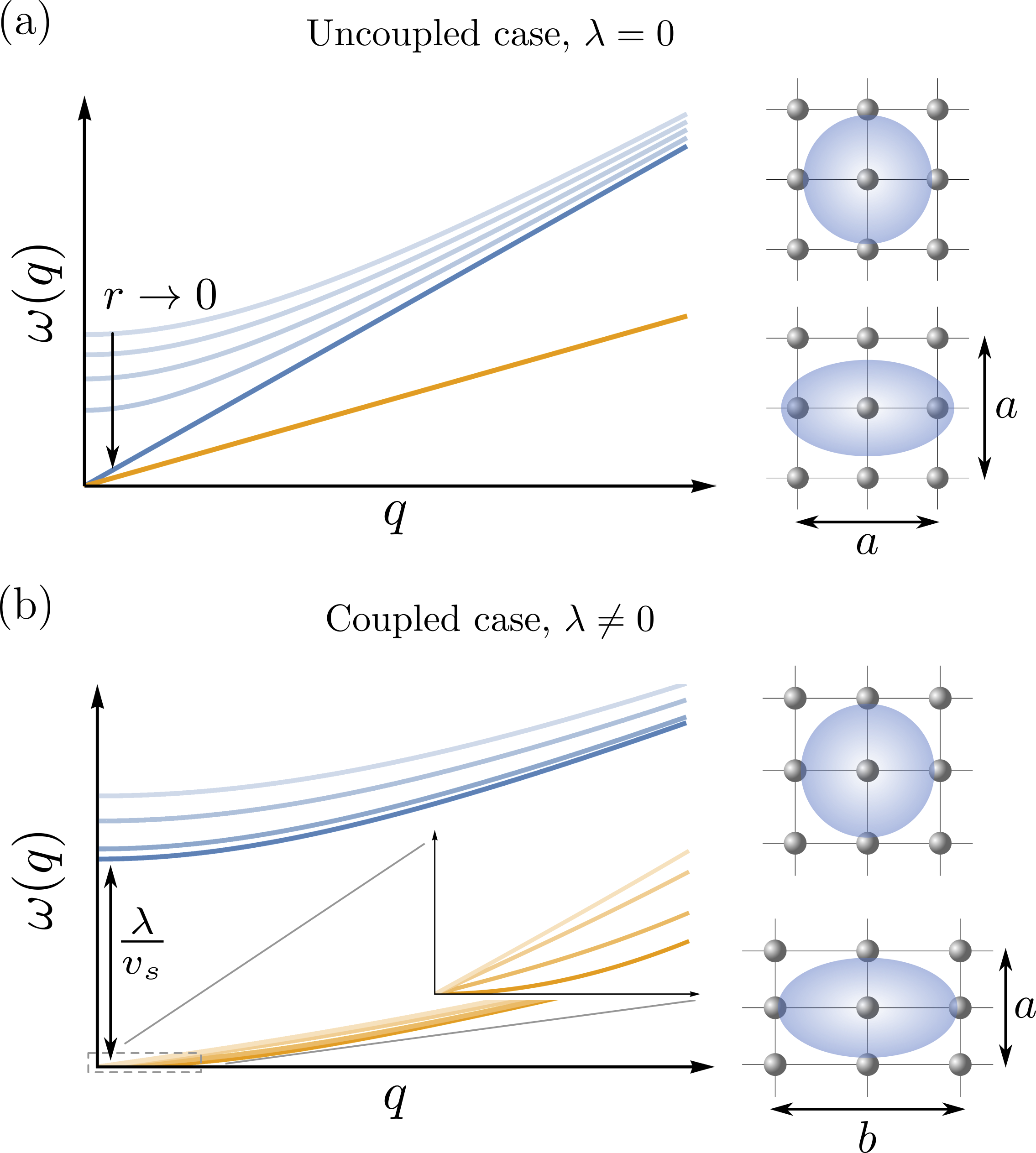}
\caption{\label{fig:adler_thrm_illustration}\textbf{Nemato-elastic modes without low-energy electronic excitations.} (a) Uncoupled nematic and acoustic phononic dispersions near an electronic nematic critical point located at $r=0$. The mass of the nematic mode reaches zero at the critical point, while the phonons are unaffected. This is illustrated schematically on the right; the Fermi surface distorts at the nematic transition while the lattice remains tetragonal. (b) Coupled nematic and phononic degrees of freedom near the electronic nematic transition. In this case, the nematic fluctuations do not go soft near the transition. Instead, the velocity of the phonon vanishes, as is expected at a structural transition, and the phonon dispersion becomes quadratic [see inset in (b)]. In this case, the Fermi surface distortion is accompanied by a tetragonal-to-orthorhombic lattice distortion.}
\end{figure}

Before considering the details of a microscopic coupling between nematic fluctuations and phonons, it is helpful to review what happens when coupling a massive field to a Goldstone mode. In our case, the nematic fluctuations, $\phi$, constitute the massive field while the phonon, associated with the lattice displacement $\mbf{u}$, is the Goldstone mode. Recall that the nematic order parameter couples to the spatial derivatives of $\mathbf{u}$ via its coupling to the strain $\varepsilon_{ij}=(\partial_i u_j + \partial_j u_i)/2$. In the tetragonal lattice and in the case of nematic order with $d_{x^2-y^2}$ symmetry, for example, the coupling is of the form $\phi (\varepsilon_{xx} - \varepsilon_{yy})$.  For simplicity, we will consider only the transverse phonons in what follows. Assuming that the two fields are real and using the nemato-elastic coupling above, the Ginzburg-Landau free energy density to quadratic order is
\begin{align}
    f &= \sum_{\mbf{q}}\begin{pmatrix}
        \phi_{-\mbf{q}} & u_{-\mbf{q}}^T
    \end{pmatrix} 
    \begin{pmatrix}
        r + v_n^2 q^2 & i \lambda q \\
        -i \lambda q & v_s^2 q^2
    \end{pmatrix}
    \begin{pmatrix}
        \phi_{\mbf{q}} \\
        u_{\mbf{q}}^T
    \end{pmatrix}\,, \label{eq:f_mat}
\end{align}
where $u_{\mbf{q}}^T$ denotes the Fourier component of the transverse eigenmode of the elasticity problem, $v_s$ is the phonon velocity, $\lambda$ is the nemato-elastic coupling constant, $r$ is the ``mass'' (i.e., frequency) of the nematic mode, and $v_n$ is the velocity of the nematic mode. Moreover, the momentum $q$ here is along the diagonal direction (see Sec.~\ref{sec:coupled_system} for a derivation of this expression). We note that we have chosen units such that $\mbf{q}$ is dimensionless. In the absence of a coupling $\lambda$ between the two fields, the nematic transition occurs at $r=0$. As this point is approached, the dispersion of the nematic mode progresses from being quadratic in $q$ away from the transition to being linear in $q$ at the critical point, as shown in Fig.~\ref{fig:adler_thrm_illustration}(a). Since the modes are decoupled, the dispersion of the phonon mode is unaffected by this and remains linear regardless of the value of $r$. 

When the two fields are coupled, the location of the nematic critical point is shifted to a finite value of $r$. This can be seen by integrating out the phonons, which yields an effective free energy density for the nematic fluctuations:
\begin{equation}
    f_{\rm eff} = \sum_q \phi_{-q}\left(r - \frac{\lambda^2}{v_s^2} + v_n^2 q^2 \right)\phi_q\,.
\end{equation}
Hence, the nematic transition occurs at $r_{\ast}=\frac{\lambda^2}{v_s^2}$. We thus define $r = \frac{\lambda^2}{v_s^2} + x$ such that $x$ measures the distance to the renormalized critical point. As the phonons are Goldstone modes, they must couple to the nematic fields \emph{via} a gradient coupling -- a direct consequence of Adler's theorem~\cite{Adler1965Consistency} -- which gives the linear-in-$q$ coupling term in Eq.~\eqref{eq:f_mat}. This has important consequences for the behavior near the critical point, as depicted in Fig.~\ref{fig:adler_thrm_illustration}(b), which shows the eigenmodes of the system as the nematic transition is approached. In particular, the velocity of the massless mode vanishes at the critical point, while the other mode remains massive. These phenomena are easily seen from expanding the modes near the transition ($x \ll \frac{\lambda^2}{v_s^2}$) for small $q^2 \ll \frac{\lambda^2}{v_s^2}\frac{1}{v_n^2 + v_s^2}$:
\begin{align}
    \omega_{-}^2(q) &= \frac{v_s^4}{\lambda^2} x q^2 + \frac{v_s^4}{\lambda^2}\left(v_n^2- \frac{v_s^4 + 2 v_n^2 v_s^2}{\lambda^2}x \right) q^4 \\
    \omega_{+}^2(q) &= \frac{\lambda^2}{v_s^2} + x + \left(v_n^2 + v_s^2 + \frac{v_s^4}{\lambda^2} x \right) q^2\,.
\end{align}
Two important and well-known results follow from this analysis: (i) The Goldstone mode remains massless even after coupling to the massive nematic fluctuations~\cite{Adler1965Consistency}. This is guaranteed by the gradient coupling in Eq.~\eqref{eq:f_mat}. (ii) There is no soft collective nematic mode at the critical point. Instead, the transition is characterized by a vanishing of the velocity of the massless mode, which acquires a quadratic dispersion~\cite{Weber2018,Merritt2020,Birgeneau2021}. This behavior is typical of structural transitions that lower the crystal symmetry~\cite{Cowley1976Acoustic}. The massive mode is separated from the Goldstone mode by a gap, $m =\sqrt{\frac{\lambda^2}{v_s^2}+x}$ which remains finite even at the critical point. In other words, the soft mode is ``eaten'' by the Goldstone mode. 

Note that the situation would be qualitatively different if the coupling in Eq.~\eqref{eq:f_mat} was not via a gradient term, i.e., if the coupling was not to a Goldstone mode. This is the case for a coupling between the nematic mode and an optical phonon mode that transforms as the same irreducible representation as the nematic mode. In this case, $\lambda q \rightarrow \lambda$ in Eq. (\ref{eq:f_mat}) and the result is that one of the hybridized modes is softened whereas the other one is hardened. Consequently, as the critical point is approached, the frequency of one of the two hybrid optical phonon/nematic modes continuously vanishes.

\section{Uncoupled nematic fluctuations and lattice vibrations}

\subsection{Nematic fluctuations}\label{sec:nematic_fluctuations}

Before discussing the electronically induced coupling between phonons and nematic fluctuations, we briefly review the dynamics of electronic nematic fluctuations in a metal and how cold spots appear in the spectrum~\cite{Oganesyan01,Garst2009,Sachdev2014}. As described in the introduction, the collective modes of a system near an electronic $d_{x^2-y^2}$ Ising-nematic transition can be represented as emergent bosonic excitations
\begin{equation}
    \phi(\mathbf{q},t) = \sum_{\mathbf{k}}\cos(2\theta_{\mbf{k}})\bar{\psi}_{\mathbf{k}+\frac{\mathbf{q}}{2}}(t)\psi_{\mathbf{k}-\frac{\mathbf{q}}{2}}(t)\,,\label{eq:nem_def}
\end{equation}
which represents the nematic field in what follows. Here, $\bar{\psi}$ and $\psi$ denote fermionic Grassman fields and the form-factor $\cos(2\theta_{\mbf{k}})$ corresponds to the breaking of fourfold rotational symmetry while preserving horizontal and vertical mirror symmetries. Note that, in the case of a $d_{xy}$ Ising-nematic, the form factor is $\sin(2\theta_{\mbf{k}})$, which preserves the diagonal mirror symmetries. $\theta_{\mbf{k}}$ is the angle between $\mbf{k}$ and the $k_x$-axis. The dynamics of the emergent field $\phi$ is inherited from the electronic quasiparticles through their minimal coupling described by the action
\begin{equation}
    \mathcal{S}[\bar{\psi},\psi,\phi] = \mathcal{S}_{\rm el}[\bar{\psi},\psi] + \mathcal{S}_{\rm nem}[\phi] + \mathcal{S}_{\rm el-nem}[\bar{\psi},\psi,\phi]\,.
\end{equation}
The electronic part describes a metallic state and is given by
\begin{equation}
    \mathcal{S}_{\rm el}[\bar{\psi},\psi] = \sum_k \bar{\psi}_k \left[ i\omega_n - \epsilon(\mbf{k}) \right] \psi_k \label{eq:el_action}
\end{equation}
where the summation index $k=(\mbf{k},\omega_n)$, $\omega_n$ is a fermionic Matsubara frequency and $\epsilon(\mbf{k})$ is the dispersion, which for simplicity we take to be parabolic, $\epsilon(\mbf{k})=\frac{\mbf{k}^2}{2m}-\mu$.
The quadratic part of the action for the nematic field is
\begin{equation}
    \mathcal{S}_{\rm nem}[\phi] = \sum_q \phi_{-q} \chi^{-1}_0(\mbf{q},\Omega_n) \phi_q\,,
\end{equation}
where again $q = (\mbf{q},\Omega_n)$, $\Omega_n$ is a bosonic Matsubara frequency, and the bare propagator takes the minimal form
\begin{equation}
    \chi^{-1}_0(\mbf{q},\Omega_n) = r + v_{n}^2 \mbf{q}^2 + \Omega_n^2 \,,
\end{equation}
where we have scaled away an unimportant overall constant. Here, $r$ denotes the distance to the bare nematic quantum critical point (QCP) and $v_{n}$ is the velocity of the nematic fluctuations. As the nematic fluctuations are electronic in origin, the dynamics of the electronic quasiparticles determines the dynamics of the emergent nematic fields. This is achieved through a Yukawa-like coupling which, in the long-wavelength, $|\mbf{q}|\rightarrow 0$, limit reads
\begin{equation}
    \mathcal{S}_{\rm el-nem}[\bar{\psi},\psi,\phi] = \tilde{\lambda}_{\rm nem} \sum_{kq} k_F^2\cos(2\theta_{\mbf{k}})  \bar{\psi}_{k+\frac{q}{2}} \psi_{k-\frac{q}{2}} \phi_q \label{eq:el_nem_coupling}\,.
\end{equation}
Note that this term is invariant under all symmetries since $\phi$ has the same symmetry as $\cos(2\theta_{\mbf{k}})$ [see Eq.~\eqref{eq:nem_def}]. Upon integrating out the electronic degrees of freedom, an effective theory for the nematic fluctuations is obtained:
\begin{equation}
    \mathcal{S}_{\rm eff-nem}[\phi] = \sum_{q}\phi_{-q} \chi_{\rm dres}^{-1}(\mbf{q},\Omega_n) \phi_q\,,
\end{equation}
with the dressed nematic propagator
\begin{equation}
    \chi_{\rm dres}^{-1}(\mbf{q},\Omega_n) =r + v_n^2 \mathbf{q}^2 + \Omega_n^2  + \Pi_{\rm nem}(\mbf{q},\Omega_n) \,. \label{eq:bare_nem_prop_matsubara}
\end{equation}
The poles of the nematic propagator on the real frequency axis determine the dispersion of the collective nematic modes. This is obtained \emph{via} analytical continuation, $i\Omega_n \rightarrow \omega + i0^+$, in which case we have
\begin{equation}
    \chi^{-1}_0(\mbf{q},\omega) = r + v_{n}^2 q^2 -\omega^2 \,,
\end{equation}
and the nematic polarization bubble is
\begin{equation}
    \Pi_{\rm nem}(\mbf{q},\omega) = - \lambda_{\rm nem}^2 \bigg[ 2 \cos^2 (2\theta_{\mbf{q}}) g(\tfrac{\omega}{v_F q}) - \cos(4\theta_{\mbf{q}}) f(\tfrac{\omega}{v_F q}) \bigg]\,.\label{eq:nem_bubble}
\end{equation}
Here we have redefined the coupling constant, $\lambda_{\rm nem}^2 \equiv \tfrac{k_F^4 N_F}{4}\tilde{\lambda}_{\rm nem}^2$, 
implying that $\lambda_{\rm nem}$ has units of energy. We note that the specific angular dependence in Eq.~\eqref{eq:nem_bubble} originates from the choice of the $\cos (2\theta_{\mbf{k}})$ form-factor although the dynamic critical exponents for modes along $\theta_{\mbf{q}}=0$ and $\theta_{\mbf{q}}=\tfrac{\pi}{4}$, respectively, do not depend on this choice for an isotropic system~\cite{Garst2009}. In the above, $v_F$ is the Fermi velocity, $\theta_{\mbf{q}}$ is the angle between $\mbf{q}$ and the $x$-axis, and
\begin{align}
    f(z) &= 1 + 4z^2 - 8 z^4 + i8 z^3 \sqrt{1-z^2} \label{eq:f_func} \\
    g(z) &= 1 + i \frac{z}{\sqrt{1-z^2}}\,. \label{eq:g_func}
\end{align}
In the static limit, $\omega = 0$, the polarization bubble is $\Pi_{\rm nem}(\mbf{q},0) = -\lambda_{\rm nem}^2$. As a consequence, the nematic QCP is shifted to $r_{\ast}=\lambda_{\rm nem}^2$. In what follows, we will therefore define $x \equiv r - \lambda_{\rm nem}^2$ such that $x$ measures the distance to the renormalized QCP. 

\begin{figure}[t]
    \centering
    \includegraphics[width=\columnwidth]{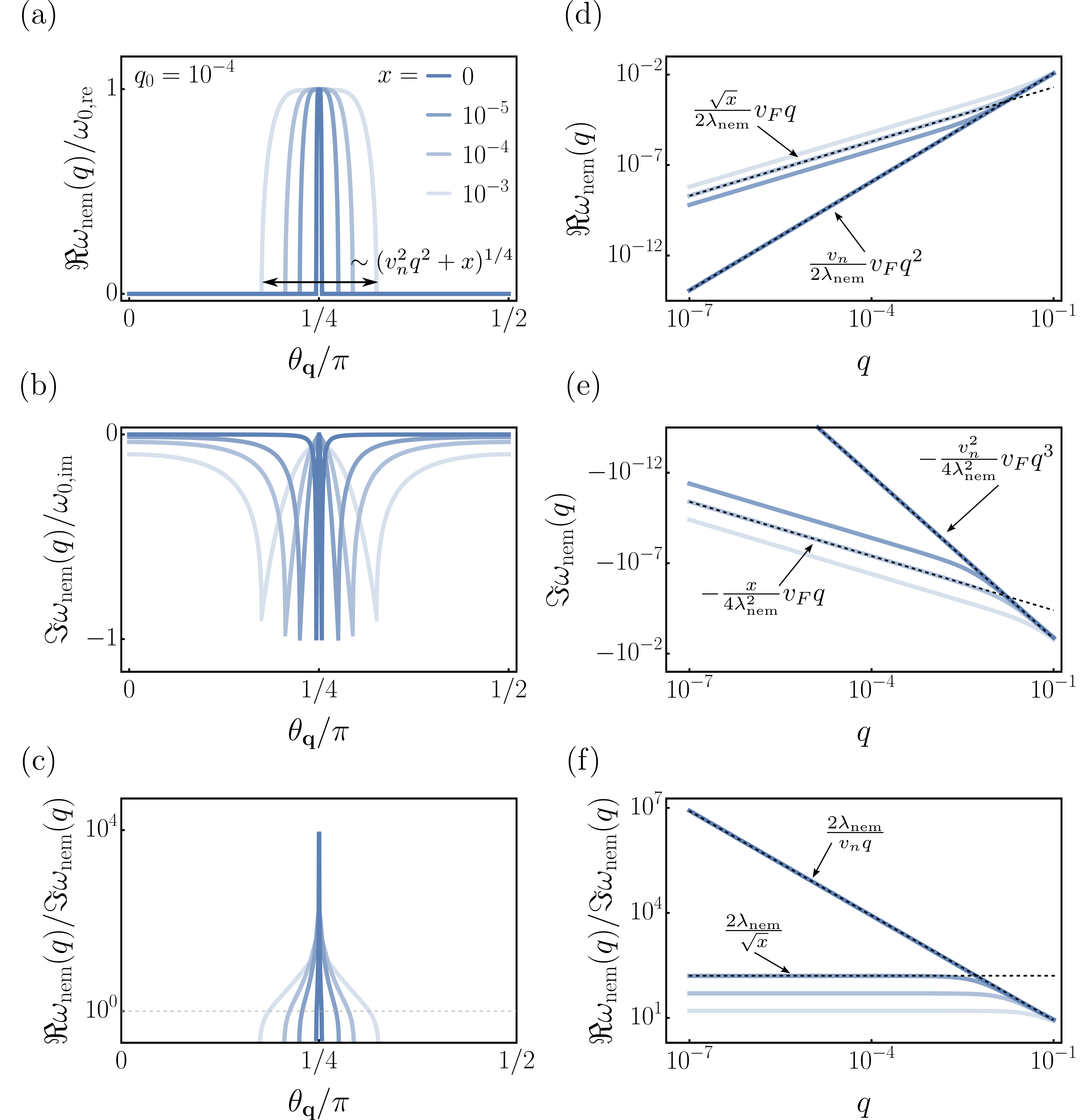}
    \caption{\label{fig:nematic_fluctuations_only}\textbf{Evolution of the nematic modes uncoupled from the lattice.} Nematic dispersion for $v_n/v_F=0.60$ and $\lambda_{\rm nem}/v_F = 0.25$ as the nematic QCP is approached. (a)--(c) Angular-dependence and (d)--(f) $q$-dependence of the dispersion at $\theta_{\mbf{q}}=\tfrac{\pi}{4}$ for different values of $x$ approaching the QCP. The real (a) and imaginary (b) parts of the angular dependence have been normalized by $\omega_{0,\mathrm{re}}$ and $\omega_{0,\mathrm{im}}$, respectively, where $\omega_{0,\mathrm{re}}$ is the value of the dispersion at $\theta_{\mbf{q}}=\tfrac{\pi}{4}$ and $q=q_0$ [see Eq.~\eqref{eq:omega_sol_3}] and $\omega_{0,\mathrm{im}}$ is the imaginary part of the dispersion at the angle where the real part vanishes and the imaginary part peaks. The angular dependence of the ratio between the real and the (absolute value of the) imaginary part is shown in (c), plotted on a log-scale. Note that the width of the region where the real part is finite scales with $(v_n^2 q^2 + x)^{1/4}$, and thus narrows with $\sqrt{q}$ at the QCP. The $q$-dependence for fixed $\theta_{\mbf{q}}=\tfrac{\pi}{4}$ in (d)--(f) is shown on a log-log scale to highlight the different regimes discussed in the text. For finite $x$, the dependence is linear in small $q$, while at the QCP the real part becomes quadratic and the imaginary part becomes cubic in $q$. Consequently, the ratio, shown in (f), tends to a constant for finite $x$ but diverges as $q\rightarrow 0$ at the QCP.}
\end{figure}

The analytic form of Eqs.~\eqref{eq:f_func} and \eqref{eq:g_func} assumes the regime $\omega < v_F q$. Because the function  $g(z)$ has a leading contribution of the form $i \omega/(v_F q)$, and since we take both $\omega$ and $q$ to zero, this implies that our analysis is restricted to an overdamped mode. The mode becomes underdamped only in a narrow angular region around $\theta_{\bf q} = \pi/4$, and this region shrinks to zero as one approaches the QCP [see Figs. \ref{fig:nematic_fluctuations_only}(a)--(c)] as first discussed in Ref.~\cite{Oganesyan01}. The behavior of the mode is overdamped, decaying with almost no oscillations, even though it is formally massless in the sense that $\omega(q\to 0) =0$, and even though its lifetime  diverges as one approaches the QCP, $x\to 0,q\to 0$. Such behavior should be contrasted with the regime $\omega > v_F q$, for which the mode is both coherent (i.e. infinitesimally damped in the clean limit) and truly massive in the sense that $\omega(q\to0)\neq 0$, with the \emph{mass} going to zero at the QCP. Henceforth, to eschew confusion, we avoid distinguishing the modes as ``massive'' and ``massless'', describing them instead as over- or underdamped.
We note that due to the branch cuts at $z=1$ in Eqs.~\eqref{eq:f_func} and \eqref{eq:g_func}, the limits $\omega < v_F q$ and $\omega > v_F q$ are separated by cusps in the propagators. The QCP is therefore most visible in the behavior of the overdamped-to-underdamped modes in the $\omega < v_F q$ limit, which is the regime we will focus on in the remainder of the paper. Note that, because these modes are caused by the coupling to the gapless particle-hole excitations of the metal, they can be thought of as hybrid nematic-plasmonic modes.

It is convenient to consider the long-wavelength limit, where $x,|\mbf{q}|^2,|\omega|^2$ are all assumed to be infinitesimally small. This is the simplest way of realizing the crossover advertised above from an overdamped to an underdamped mode as $x\rightarrow 0$. In this limit, the modes are (at least) linear in $|\mbf{q}|$ with a coefficient that is much smaller than $v_F$ implying that 
\begin{equation}
    z=\frac{\omega}{v_F |\mbf{q}|} \ll 1\,.\label{eq:z-def}
\end{equation}
As explained above, this is the relevant regime for exploring the quantum dynamics of the system. The opposite limit is explored in more detail in Appendix~\ref{app:nematic_modes_details}.

Expanding to second order in $z$ we obtain the dressed nematic susceptibility,
\begin{equation}
    \chi^{-1}_{\mathrm{dres}}(\mbf{q},\omega)_{(2)}
    = \lambda_{\rm nem}^2\left(y - 2i \cos^2 2\theta_{\mbf{q}} z + 4  \cos 4 \theta_{\mbf{q}} z^2\right)\,, \label{eq:nem_susc_2nd_order-new}
\end{equation}
where 
\begin{equation}
    y = \frac{x + v_n^2|\mbf{q}|^2}{\lambda_{\rm nem}^2} \ll 1.
\end{equation}
Focusing on the pole with the negative imaginary part, corresponding to the physical solution, we find
\begin{equation}
    \omega_{(2)}(|\mbf{q}|) = v_F|\mbf{q}|\frac{\cos^2 2\theta_{\mbf{q}}}{4\cos 4\theta_{\mbf{q}}} i \left[1 - \sqrt{1+\frac{4 y \cos 4\theta_{\mbf{q}}}{\cos^4 2\theta_{\mbf{q}}}}\right]
\end{equation}
which has a real part only if
\begin{equation}
    \cos^2(2\theta_{\mbf{q}}) < 2\sqrt{y},
\end{equation}
where again we expanded in small $y$. Thus, for
\begin{equation}
    |\theta_{\mbf{q}}-\pi/4|>(y/4)^{1/4},
\end{equation}
the mode is a purely damped one, with only an imaginary component.

Along the diagonal, $\theta_\mbf{q} = \pi/4$, the linear term in Eq.~\eqref{eq:nem_susc_2nd_order-new} vanishes. To understand the behavior of the mode along this direction, we expand the susceptibility to third order,
\begin{equation}
    \chi^{-1}_{\mathrm{dres}}(\mbf{q},\omega)_{(3)}=\chi^{-1}_{\mathrm{dres}}(\mbf{q},\omega)_{(2)}-8i \lambda_{\rm nem}^2z^3, \label{eq:eq:nem_susc_3rd_order-new}
\end{equation}
from which we obtain
\begin{equation}
    \omega_{(3)}(|\mbf{q}|) = v_F|\mbf{q}|\sqrt{\frac{y}{4}}\left(1-i\sqrt{\frac{y}{4}}\right). \label{eq:omega_sol_3}
\end{equation}
This makes it clear that there are two different regimes of interest, characterized by whether $v_n^2 q^2 \ll x$ or $v_n^2 q^2 \gg x$. In the former case, the long-wavelength behavior is dominated by a linear-in-$q$ term:
\begin{equation}
    \omega(q) \approx \left(\frac{\sqrt{x}}{2\lambda_{\rm nem}} - i \frac{x}{4\lambda_{\rm nem}^2} \right) v_F q \label{eq:omega_nem_linear_q_no_ph}
\end{equation}
which characterizes the behavior of non-critical fluctuations far from the critical point. In the opposite limit, the long-wavelength behavior is is characterized by
\begin{equation}
    \omega(q) \approx \frac{v_F v_n}{2\lambda_{\rm nem}} q^2 - i \frac{v_F v_n^2}{4\lambda_{\rm nem}^2}q^3\,, \label{eq:omega_nem_higher_q_no_ph}
\end{equation}
which is the relevant term near the QCP. The crossover between the two regimes occurs at $q \sim \frac{\sqrt{x}}{v_n}$ as shown in Fig.~\ref{fig:nematic_fluctuations_only} for different values of $x$. Moreover, the behavior outlined above demonstrates that the inverse damping -- proportional to the scattering time -- defined by the ratio between the real and the (absolute value of the) imaginary part, is qualitatively different near the QCP. In the region where $v_n^2 q^2 \ll x$, the inverse damping goes as $2\lambda_{\rm nem} / \sqrt{x}$, independent of $q$. In contrast, at the QCP, the inverse damping diverges as $q \rightarrow 0$ and is given by $ 2\lambda_{\rm nem}/ v_n q$.

In Fig.~\ref{fig:nematic_fluctuations_only} we plot the roots of Eq.~\eqref{eq:bare_nem_prop_matsubara} (after analytic continuation) for specific values of $q$ along $\theta_{\mbf{q}}$ and as a function of $q$ for $\theta_{\mbf{q}}=\tfrac{\pi}{4}$.
The angular dependence of the nematic dispersion shown in Figs.~\ref{fig:nematic_fluctuations_only}(a)--(c) demonstrates that, as the nematic QCP is approached, the nematic fluctuations become more and more focused along the $\theta_{\mbf{q}}=\tfrac{\pi}{4}$-direction. This is particularly clear when considering the ratio between the real and imaginary parts shown on a log-scale in Fig.~\ref{fig:nematic_fluctuations_only}(c). Away from the nematic QCP, fluctuations in the vicinity of $\theta_{\mbf{q}}=\tfrac{\pi}{4}$ are underdamped. As the QCP is approached, this range shrinks until only fluctuations along $\theta_{\mbf{q}}=\tfrac{\pi}{4}$ are underdamped, with the remaining directions being overdamped. 

The $q$-dependence along the $\theta_{\mathbf{q}}=\tfrac{\pi}{4}$ direction is shown in Figs.~\ref{fig:nematic_fluctuations_only}(d)-(f) on a log-log scale. At the QCP, the real part of the dispersion is quadratic in $q$, while the imaginary part is cubic leading to a divergence in their ratio, $2\lambda_{\rm nem} / v_n q$, as $q \rightarrow 0$, thus signaling the emergence of a coherent mode. Away from the QCP, at finite $x$, both real and imaginary parts are linear in $q$ and the ratio tends to a constant, $2\lambda_{\rm nem}/\sqrt{x}$, as $q\rightarrow 0$, implying that the mode becomes increasingly underdamped as the QCP is approached.

\subsection{Structural transitions and transverse phonons}\label{sec:transverse_ph}

Structural transitions are characterized by the vanishing of specific (combinations of) elastic constants. For concreteness, we are here interested in the tetragonal-to-orthorhombic transition that  preserves  the vertical mirrors but breaks the diagonal mirrors [consistent with the choice of form factor made in Eq.~\eqref{eq:nem_def}]. In this case, the combination of elastic constants $c_{11}-c_{12} \rightarrow 0$. The vibrational eigenmodes near this transition are obtained from the dynamical matrix \cite{Landau_Elasticity,Carvalho2019}
\begin{widetext}
\begin{equation}
    M_{ij}(\mbf{q}) = \begin{pmatrix}
        c_{11}q_x^2 + c_{66}q_y^2 + c_{44} q_z^2 & (c_{12}+c_{66})q_x q_y & (c_{13}+c_{44})q_x q_z \\
        (c_{12}+c_{66})q_x q_y & c_{66}q_x^2 + c_{11}q_y^2 + c_{44} q_z^2 & (c_{13}+c_{44})q_y q_z \\
        (c_{13}+c_{44})q_x q_z & (c_{13}+c_{44})q_y q_z & c_{44}(q_x^2 + q_y^2) + c_{33} q_z^2 
    \end{pmatrix}\,. \label{eq:Mij}
\end{equation}
\end{widetext}
We note that, since the matrix of elastic constants is $6 \times 6$, the eigenvectors of the $3 \times 3$ dynamical matrix do not generally produce the combinations of strains that are eigenvectors of the elastic constant matrix~\cite{Cowley1976Acoustic}. As $c_{11}-c_{12} \rightarrow 0$, the velocity of the modes propagating along $[110]$ and $[1\bar{1}0]$ vanishes. Importantly, these are transverse modes with polarization perpendicular to the direction of propagation. Hence, this tetragonal-to-orthorhombic structural transition is associated with a vanishing velocity of transverse phonons propagating along a diagonal direction. 

Describing how nematic fluctuations are affected by soft lattice vibrations thus requires coupling the electrons and transverse phonons. In the simplest case, the electron-phonon matrix element is proportional to $\mbf{e}_{\mbf{q},\alpha} \cdot \mbf{q}$, where $\mbf{e}_{\mbf{q},\alpha}$ is the phonon polarization vector, $\alpha$ labels the polarization, and $\mbf{q}$ is the momentum. Hence, at lowest order in $q$, the coupling between electrons and transverse phonons vanishes. This can be remedied by considering the effects of, e.g., electronic orbitals~\cite{Gastiasoro2023Generalized} or impurities~\cite{Schmid1973Electron-phonon}. Here, we will pursue the latter approach which is based on how the scattering potential of impurities is altered as the lattice distorts. Writing the coupling between electrons and the phonon displacement field $u_{q,\alpha}$ as
\begin{equation}
    \mathcal{S}_{\rm el-ph} =i \tilde{\lambda}_{\rm ph} \sum_{\mbf{k}\mbf{k'}\alpha} M_{\alpha}(\mbf{k},\mbf{k'})\bar{\psi}_{\mbf{k}} \psi_{\mbf{k'}} u_{\mbf{k}-\mbf{k'},\alpha}\,,\label{eq:el_ph_coupling}
\end{equation}
the electron-phonon matrix element is found to be~\cite{Tsuneto1961Ultrasound,Schmid1973Electron-phonon}
\begin{equation}
    M_{\alpha}(\mbf{k},\mbf{k}') \propto \left(\mbf{p}\cdot\mbf{q}\right)\left( \mbf{e}_{\mbf{k}-\mbf{k}',\alpha} \cdot \mbf{p} \right)\big|_{\substack{\mbf{q}=\mbf{k}-\mbf{k}' \\ \mbf{p}=\frac{1}{2}\left( \mbf{k}+\mbf{k}'\right)}}\,.
\end{equation}
For transverse phonons $u^{T}_{q}$ we thus find
\begin{equation}
    M_{\mbf{k}+\mbf{q}/2,\mbf{k}-\mbf{q}/2} = k^2 q \cos \left(\theta_{\mbf{k}}-\theta_{\mbf{q}} \right) \sin \left( \theta_{\mbf{k}}-\theta_{\mbf{q}} \right)\,, \label{eq:el_TA_coupling}
\end{equation}
where $\theta_{\mbf{k}}$ ($\theta_{\mbf{q}}$) is the angle between $\mbf{k}$ ($\mbf{q}$) and the $x$-axis. While the matrix element only depends on the relative angle between $\mbf{k}$ and $\mbf{q}$, it is convenient to keep the explicit dependence on both angles since the nematic propagator only depends on $\theta_{\mbf{q}}$. 

The transverse phonons are described by the action
\begin{equation}
    \mathcal{S}_{\rm ph} = \sum_q u^{T}_q \mathcal{D}_0^{-1}(\mbf{q},\Omega_n) u^{T}_{-q}\,, \label{eq:kinetic_ph}
\end{equation}
with
\begin{equation}
    \mathcal{D}_0^{-1}(\mbf{q},\Omega_n) = v_{s}^2 q^2 + \Omega_n^2\,,\label{eq:bare_ph_prop}
\end{equation}
where we have once again scaled away an unimportant overall prefactor. In deriving these phonon propagators from Eq.~\eqref{eq:Mij}, we made the simplifying assumption that $c_{11} - c_{12} = 2 c_{66}$, which guarantees that the phonon dispersions are isotropic and purely transverse or purely longitudinal along all directions. The lattice anisotropy is still encoded in selecting just the $d_{x^2-y^2}$ nematic instability. In this case, anisotropic phonon modes are achieved through the coupling between electrons and TA phonons in Eq.~\eqref{eq:el_TA_coupling}. Integrating out the electrons, given by Eq.~\eqref{eq:el_action}, we find an effective action for the transverse phonons
\begin{equation}
    \mathcal{S}_{\rm eff-ph} = \sum_{q} u^{T}_q \mathcal{D}_{\rm dres}^{-1}(\mbf{q},\Omega_n) u^{T}_{-q}\,.
\end{equation}
Here, the dressed phonon propagator in real frequencies is
\begin{equation}
    \mathcal{D}^{-1}_{\rm dres}(\mbf{q},\omega) =  v_s^2 q^2 - \omega^2 + \Pi_{\rm ph}(\mbf{q},\omega) \label{eq:dressed_phonon}
\end{equation}
where the transverse phonon polarization bubble is
\begin{equation}
    \Pi_{\rm ph}(\mbf{q},\omega) = - \lambda_{\rm ph}^2 q^2 f\left( \tfrac{\omega}{v_F q}\right)\,,\label{eq:ph_bubble}
\end{equation}
with the coupling constant $\lambda_{\rm ph}^2 \equiv \tfrac{k_F^4 N_F}{16}\tilde{\lambda}_{\rm ph}^2$ and $f(\tfrac{\omega}{v_F q})$ defined in Eq.~\eqref{eq:f_func}. Consequently, the coupling to the electrons softens the transverse phonons by reducing the sound velocity, although the phonons remain perfectly coherent with vanishing imaginary part.

At this point we remark that structural transitions are typically split into two scenarios. One is akin to the traditional Peierls instability, in which the electrons drive a softening of the lattice vibrations, i.e., the vanishing of the sound velocity is primarily caused by the electron-phonon coupling. In this scenario the structural transition is electronically driven. In the other scenario, the primary driving force behind the soft lattice vibrations are the phonons themselves and the vanishing of the sound velocity is primarily due to the bare elastic constants going to zero. This is the structurally driven scenario. While such a split is only possible at a theoretical level (see, e.g., Ref.~\cite{Johannes2008Fermi}), it permits a distinction between electronically and structurally dominated transitions. In the remainder of this work, we will focus on the electronically dominated scenario and assume that, absent a coupling to the electrons, the phonon velocity would remain finite and no structural transition would occur. With this, we proceed to study the impact of including the electronically mediated coupling between transverse phonons and nematic fluctuations.

\section{Coupled nematic-phononic system}\label{sec:coupled_system}

From the considerations presented above, we draw the following conclusions, which will be important for the discussions to come:

\begin{itemize}
    \item No additional soft mode appears at the nematic QCP once these are coupled to the transverse phonons. Instead, the phonon velocity vanishes at the nematic QCP.
    \item The nematic fluctuations are underdamped in a region near $\theta_{\mbf{q}}=\tfrac{\pi}{4}$, outside of which they are overdamped.
    \item The $q$-dependence of the damping changes near the nematic QCP.
\end{itemize}
As we show below, although the details change, the phenomenology of the above results remain in place.

\subsection{The hybridized propagator}

The action of the coupled system reads
\begin{equation}
    \mathcal{S} = \mathcal{S}_{\rm el}+\mathcal{S}_{\rm ph}+\mathcal{S}_{\rm nem}+\mathcal{S}_{\rm el-ph}+\mathcal{S}_{\rm el-nem}\,,
\end{equation}
with the individual terms defined in Eqs.~\eqref{eq:el_action}--\eqref{eq:el_nem_coupling}, Eq.~\eqref{eq:el_ph_coupling}, and Eq.~\eqref{eq:kinetic_ph}. Note that, in contrast to previous phenomenological approaches \cite{Karahasanovic16,Paul17,Carvalho2019}, there is no direct coupling between the nematic fluctuations and the transverse phonons. Instead, the effective coupling between the two is mediated by the electronic degrees of freedom. Integrating out the electrons, we find that the effective action of the coupled phonon-nematic system is
\begin{align}
    \mathcal{S}_{\rm eff} = \sum_q \begin{pmatrix}
        u^{T}_q & \phi_q
    \end{pmatrix}
    \underline{\underline{\mathcal{K}}}^{-1}(q)
    \begin{pmatrix}
        u^{T}_{-q} \\
        \phi_{-q}
    \end{pmatrix}\,,
\end{align}
where the matrix $\underline{\underline{\mathcal{K}}}^{-1}(q)$, written as a function of real frequencies, is
\begin{align}
        \underline{\underline{\mathcal{K}}}^{-1}= \begin{pmatrix}
            \mathcal{D}^{-1}_{\rm dres}(\mbf{q},\omega) & \Pi_{\rm ph-nem}(\mbf{q},\omega) \\
            -\Pi_{\rm ph-nem}(\mbf{q},\omega) & \chi^{-1}_{\rm dres}(\mbf{q},\omega)
        \end{pmatrix}\,. \label{eq:prop_matrix}
\end{align}
The nematic and phononic polarization bubbles are given in Eqs.~\eqref{eq:nem_bubble} and \eqref{eq:ph_bubble}, respectively, while
\begin{equation}
    \Pi_{\rm ph-nem}(\mbf{q},\omega) = -i \lambda_{\rm nem} \lambda_{\rm ph} q \sin (2\theta_{\mbf{q}}) f\left( \tfrac{\omega}{v_F q} \right)\,, \label{eq:nem_ph_bubble}
\end{equation}
where the function $f\left( \tfrac{\omega}{v_F q} \right)$ is given in Eq.~\eqref{eq:f_func}. We note that the displacement field, $u^{T}_q$, is purely imaginary in our case.

The polarization bubble, $\Pi_{\rm ph-nem}(\mathbf{q},\omega)$ manifests the effective coupling between the nematic fluctuations and the transverse phonons and implies that the eigenmodes of the system are no longer pure nematic and pure phononic, but are instead mixtures of the two. Nevertheless, the singular behavior of the low-energy model is still governed by the zeros of $\underline{\underline{\mathcal{K}}}^{-1}$. Hence, we identify the diagonal element $(\underline{\underline{\mathcal{K}}})_{11}$ as the renormalized ``phonon'' propagator:
\begin{equation}
    \mathcal{D}^{-1}(\mbf{q},\omega) = \frac{\det\left(\underline{\underline{\mathcal{K}}}^{-1} \right)}{\chi_{\rm dres}^{-1}(\mbf{q},\omega)}\,. \label{eq:full_ph_prop}
\end{equation}
Conversely, we identify the diagonal element $(\underline{\underline{\mathcal{K}}})_{22}$ as the renormalized propagator of the ``nematic fluctuations'':
\begin{equation}
    \chi^{-1}(\mbf{q},\omega) = \frac{\det\left(\underline{\underline{\mathcal{K}}}^{-1} \right)}{\mathcal{D}^{-1}_{\rm dres}(\mbf{q},\omega)}\,.\label{eq:full_nem_prop}
\end{equation}
Here, the quotes are used to reflect the fact that the actual collective modes are superpositions of both phononic and nematic degrees of freedom. Nevertheless, these response functions would be probed if appropriate conjugate fields were applied.

\begin{table}[!t]
    \centering
    \begin{tabular}{C{0.2\columnwidth}C{0.2\columnwidth}C{0.2\columnwidth}C{0.2\columnwidth}}
    \toprule
         $v_{F}$ & $v_n$ & $v_s$ & $\lambda_{\rm nem}$ \\
         \hline
         1 & 0.6 & 0.01 & 0.25 \\
    \bottomrule
    \end{tabular}
    \caption{\label{tab:parameters}Table of parameters kept fixed in the numerical calculations unless otherwise specified.}
\end{table}

The coupling between electrons, phonons, and nematic fluctuations impacts the inverse correlation length $\xi^{-2}$, analogously defined as $\lim_{q \rightarrow 0}\lim_{\omega \rightarrow 0}\chi^{-1}(\mbf{q},\omega)/v_n^2$, which becomes a function of angle:
\begin{equation}
    \xi^{-2} = \xi_0^{-2} - \frac{\lambda_{\rm nem}^2}{v_n^2}\left(1 + \frac{\lambda_{\rm ph}^2 \sin^2 2\theta_{\mbf{q}}}{v_{s}^2 - \lambda_{\rm ph}^2}\right)\,,
\end{equation}
where $\xi^{-2}_0= r/v_n^2$ is the bare correlation length. Thus, a divergence occurs along the diagonals, $\theta_{\mbf{q}} = \tfrac{\pi}{4} \bmod \tfrac{\pi}{2}$. Consequently, the nematic QCP occurs at
\begin{equation}
    r_{\ast} = \lambda_{\rm nem}^2\left(1 + \frac{\lambda_{\rm ph}^2 }{v_{s}^2 - \lambda_{\rm ph}^2}\right) \label{eq:nem_qcp}
\end{equation}
for modes along the diagonal~\cite{Paul17}. This is expected since this is the direction along which the soft phonons signaling a tetragonal-to-orthorhombic structural transition propagate, as discussed in Sec.~\ref{sec:transverse_ph}. Consequently, the nematic transition occurs through a divergence of the susceptibility along this direction, while the modes that are not along $\theta_{\mbf{q}}=\tfrac{\pi}{4}$ remain gapped. In what follows, we will thus let $r \rightarrow r_{\ast} + x$ where $x$ measures the distance to the renormalized QCP. Moreover, we define the effective coupling
\begin{equation}
    \lambda_{\rm p-n}^2= \frac{\lambda_{\rm ph}^2 \lambda_{\rm nem}^2}{v_{s}^2 - \lambda_{\rm ph}^2}\,,
\end{equation}
denoting the impact of the coupling between nematic and phononic degrees of freedom.

Close to $r_{\ast}$, i.e., for $x \ll \lambda_{\rm p-n}^2$, the static part of the renormalized phononic propagator in the long-wavelength limit reads
\begin{align}
    \frac{\mathcal{D}^{-1}(\mbf{q},0)}{\lambda_{\rm nem}^2 \lambda_{\rm ph}^2} &= 
    \lambda_{\rm p-n}^2\cos^2 (2\theta_{\mbf{q}}) q^2 + \frac{v_n^2}{\lambda_{\rm p-n}^4} \sin^2 (2\theta_{\mbf{q}}) q^4 \nonumber \\
    & + \frac{x}{\lambda_{\rm p-n}^4}\sin^2 (2\theta_{\mbf{q}}) q^2  + \mathcal{O}(x q^4)
\end{align}
Evidently, the phonon velocity vanishes at the critical point, but only along $\theta_{\mbf{q}}=\tfrac{\pi}{4}$. Away from that direction, the velocity remains finite. This is the expected softening of the transverse phonon mode at the structural transition. In this case, the structural transition is caused by critical nematic fluctuations,
but the softening of the transverse mode still occurs and no additional critical mode appears, similar to the simpler case discussed in Sec.~\ref{sec:adler}.

\subsection{The spectral functions}\label{sec:spectral_functions}

\begin{figure}[!t]
    \centering
    \includegraphics[width=\columnwidth]{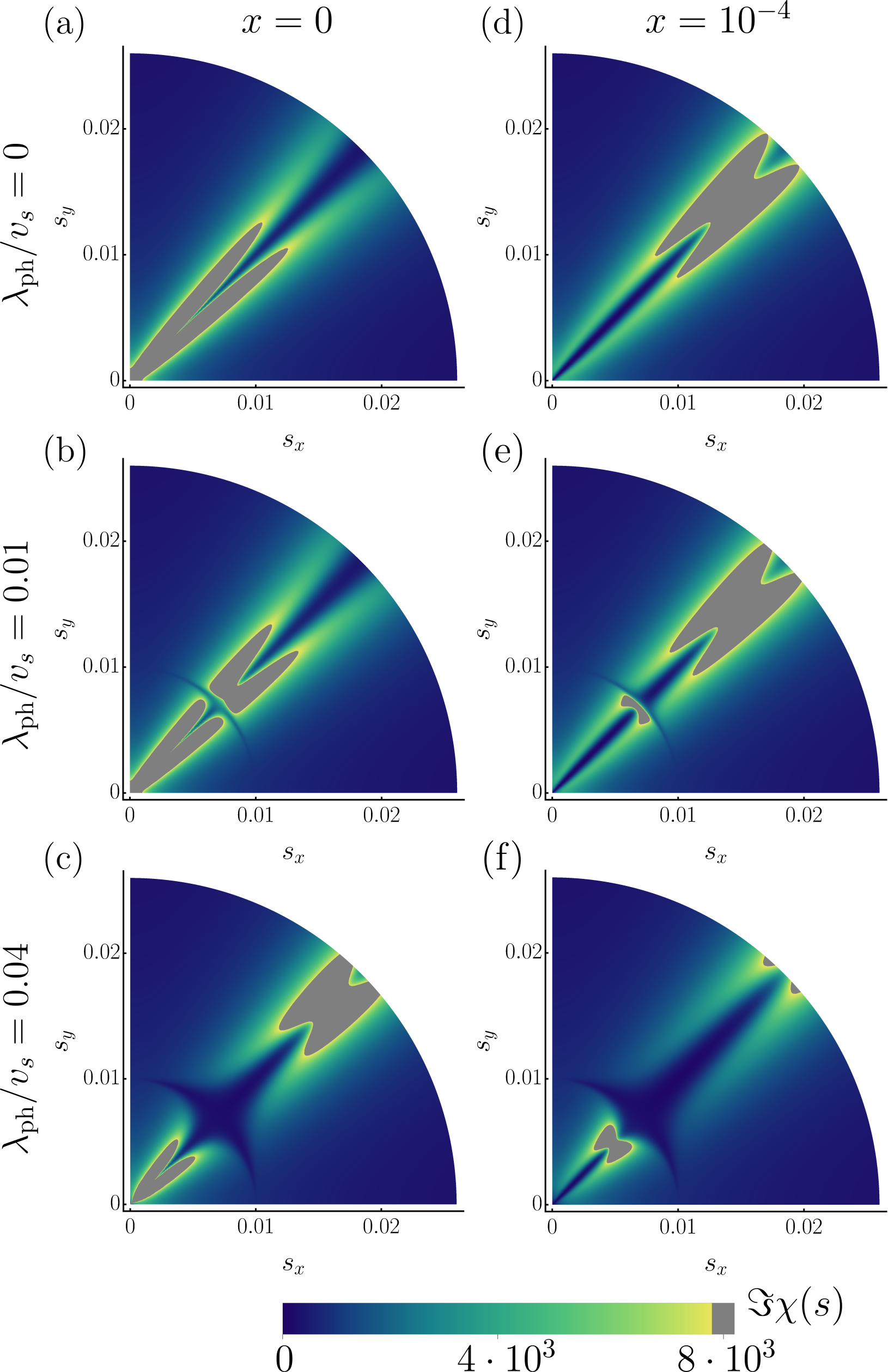}
    \caption{\label{fig:imaginary_nematic}\textbf{Imaginary part of the renormalized nematic propagator.} (a)--(c) depicts the nematic spectral function, $\Im \chi(s)$, for $x=0$ while (d)--(f) corresponds to the case $x=10^{-4}$ as a function of $s= \omega / q_0$ with $q_0 = 10^{-4}$ for $\lambda_{\rm ph}/v_s=0$ [(a) and (d)], $\lambda_{\rm ph}/v_s=0.01$ [(b) and (e)], and $\lambda_{\rm ph}/v_s=0.04$ [(c) and (f)]. As $\lambda_{\rm ph}/v_s$ becomes finite, the single nematic mode splits into two which are separated by a near-complete suppression of spectral weight.} 
\end{figure}

\begin{figure}[!t]
    \centering
    \includegraphics[width=\columnwidth]{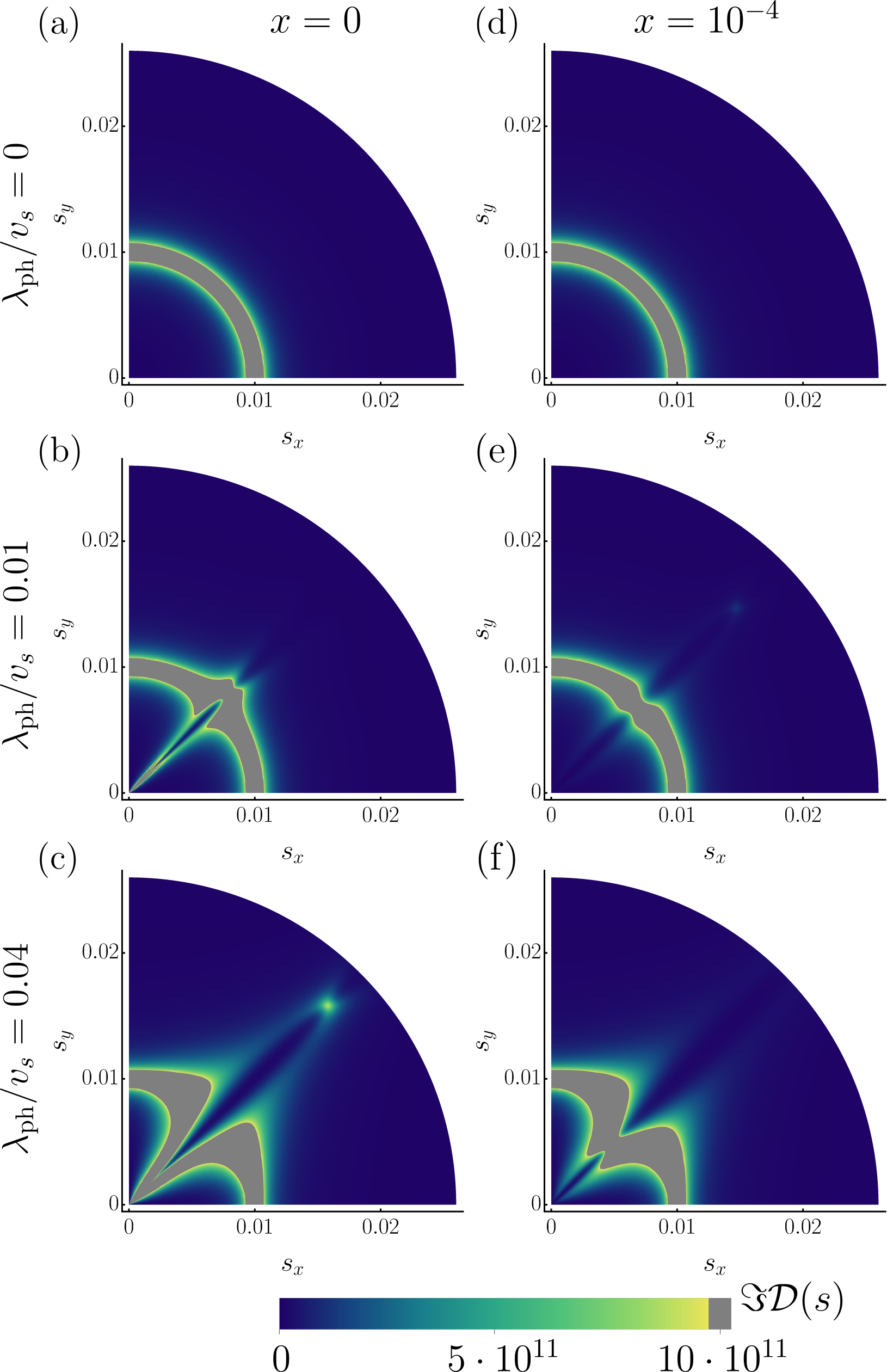}
    \caption{\label{fig:imaginary_phonon}\textbf{Imaginary part of the renormalized phonon propagator.} (a)--(c) depicts the phonon spectral function, $\Im \mathcal{D}(s)$, for $x=0$ while (d)--(f) corresponds to the case $x=10^{-4}$ as a function of $s= \omega / q_0$ with $q_0 = 10^{-4}$ for $\lambda_{\rm ph}/v_s=0$ [(a) and (d)], $\lambda_{\rm ph}/v_s=0.01$ [(b) and (e)], and $\lambda_{\rm ph}/v_s=0.04$ [(c) and (f)]. As $\lambda_{\rm ph}/v_s$ becomes finite, the phonon mode splits into two different modes, which move away from each other as $\lambda_{\rm ph}/v_s$ is increased further.} 
\end{figure}

The qualitative impact of the coupling between nematic fluctuations and phonons can be seen by considering the imaginary parts of the renormalized propagators, $\Im \chi(\mbf{q},\omega)$ and $ \Im\mathcal{D}(s)$. For this purpose, we choose a constant $q=q_0$ and $\omega = s q_0$, such that the propagators only depend on $s$. In Figs.~\ref{fig:imaginary_nematic} and \ref{fig:imaginary_phonon} we plot $\Im \chi(\mbf{q},\omega)$ and $ \Im\mathcal{D}(s)$ as functions of both $s_x$ and $s_y$ in a radial plot. The constant parameters used in this plot are given in Table~\ref{tab:parameters}, where the Fermi velocity is set equal to one. Recall that, in the $\lambda_{\rm ph}/v_s=0$ case, the phonon mode emerges at $s = v_s$ and that $x$ controls the energy of the nematic mode.

Coupling the nematic and phononic degrees of freedom has several important consequences, which are clear when comparing the panels of Figs.~\ref{fig:imaginary_nematic} and \ref{fig:imaginary_phonon}. A finite $\lambda_{\rm ph}$ blends the nematic and phononic propagators, such that two peaks appear in each spectral function, one originating from the former nematic mode and the other from the former phonon mode. Note that the second peak in the phonon propagator is nearly washed out due to the sharpness of the phonon mode, see Fig.~\ref{fig:imaginary_phonon}(e). In the nematic propagator, Fig.~\ref{fig:imaginary_nematic}, the two modes are separated by a near-complete suppression of spectral weight, consistent with the location of the original phonon mode, Fig.~\ref{fig:imaginary_phonon}(a). As $\lambda_{\rm ph}/v_s$ is increased, the two modes move further apart; however, the evolution of these modes with $\lambda_{\rm ph}/v_s$ depends strongly on $x$. To see this, consider the nematic mode in the uncoupled case, corresponding to Figs.~\ref{fig:imaginary_nematic}(a) and (d). For $x=0$, the nematic mode is at a lower value of $s$ than the phonon mode. For $x=10^{-4}$, the nematic mode moves above the phonon mode. The hybridization between the phonon and nematic modes naturally results in avoided crossings, so that the softer mode is driven even softer by the coupling, while the harder mode hardens further. 

\begin{figure}
    \centering
    \includegraphics[width=\columnwidth]{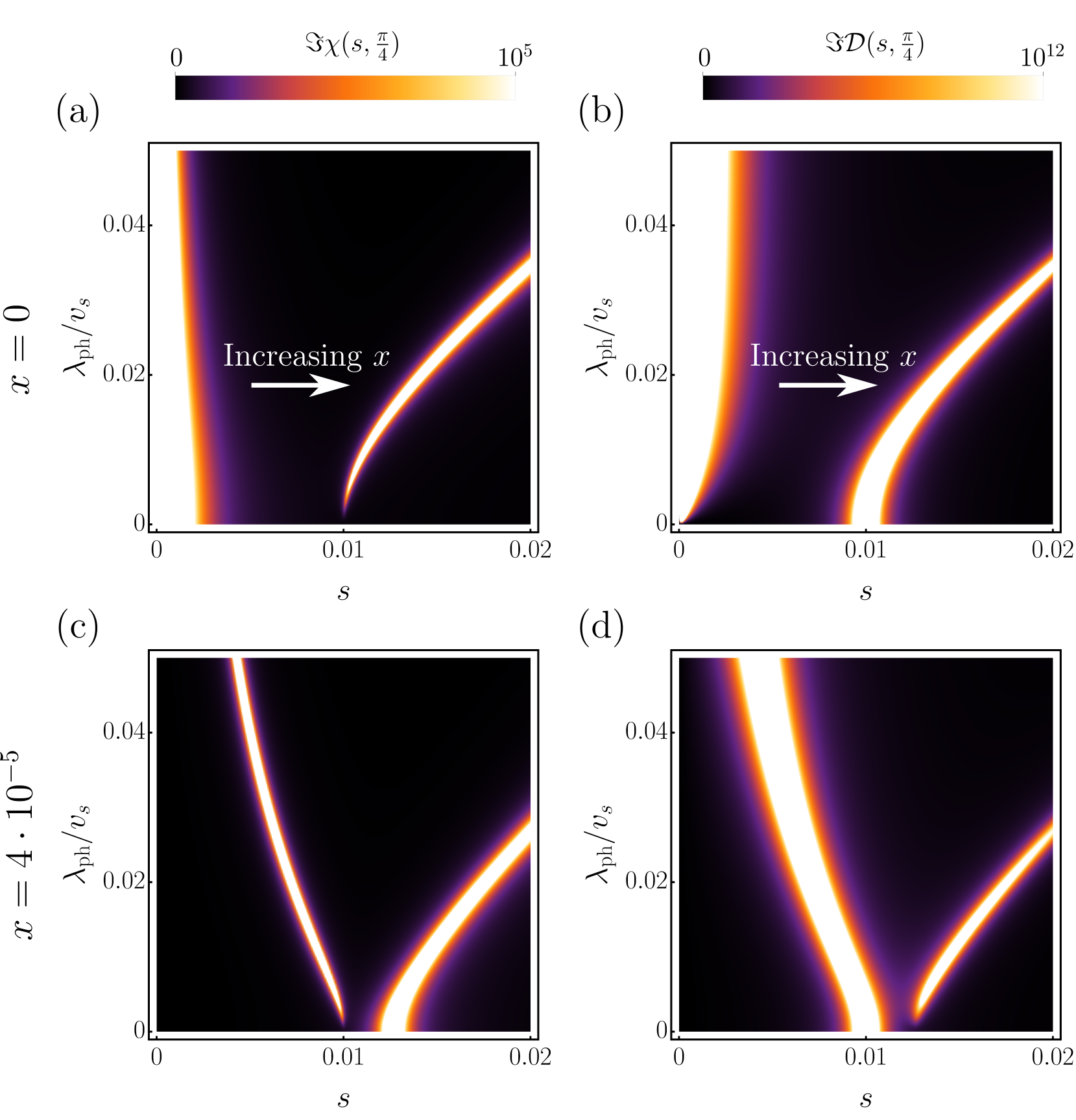}
    \caption{\label{fig:im_dispersions} \textbf{Imaginary parts of the propagators as functions of $\lambda_{\rm ph}$.} (a) and (c) show the nematic propagator while (b) and (d) show the phononic propagator (a)--(b) for $x=0$ and (c)--(d) $x=4\cdot 10^{-5}$. Here, $\theta_{\mbf{q}}=\tfrac{\pi}{4}$ and the propagators are plotted as functions of $s=\omega/q_0$ (with $q_0=10^{-4}$) and $\lambda_{\rm ph}/v_s$. As $\lambda_{\rm ph}/v_s$ is increased from 0, the lower-lying nematic mode moves to somewhat lower values of $s$ while the phonon mode frequency increases sharply.
    Note that, in (a), the phononic mode is absent at $\lambda_{\rm ph}=0$ while in (b) the nematic mode is absent at $\lambda_{\rm ph}=0$, since the two systems are uncoupled in that case. As $x$ is increased, the nematic mode moves to higher values of $s$, while the phonon mode remains at $v_s/v_F=0.01$ regardless of the value of $x$. For sufficiently large $x \gtrsim 3 \cdot 10^{-5}$, the phonon mode lies below the nematic one and consequently, the phonon mode softens upon increasing $\lambda_{\rm ph}/v_s$.}
\end{figure}

This can be more clearly seen in Fig.~\ref{fig:im_dispersions}, which depicts how the modes in both $\chi(s)$ and $\mathcal{D}(s)$ disperse with increasing coupling. As $x$ is increased, the nematic mode moves to higher values of $s$ while the phonon mode remains at $s_0=v_s q_0$, as shown in Figs.~\ref{fig:im_dispersions}(a) and (b). Once $x$ is above a threshold value which depends on $q_0$, the nematic mode reaches $s_0$, the phonon mode moves below the nematic one [Figs.~\ref{fig:im_dispersions}(c) and (d)] and the behavior of the two modes as a function of $\lambda_{\rm ph}/v_s$ changes qualitatively. We find that, near the point where the two modes would cross in the complex plane if they were uncoupled, the impact of $\lambda_{\rm ph}$ is particularly pronounced resulting in substantial mode-mode coupling, which we discuss in further detail in Appendix~\ref{app:mode_mode}.

\subsection{Dispersion relations}\label{sec:on-shell}

As the lattice is coupled to nematic fluctuations, the distinction between a phononic and a nematic mode is blurred, and the actual eigenmodes of the system are mixtures of the two. In what follows, we will therefore denote the on-shell nematic mode by the one which, in the $\lambda_{\rm ph} \rightarrow 0$ limit, reduces to the nematic mode discussed in Sec.~\ref{sec:nematic_fluctuations}. Similarly, the on-shell phonon mode is characterized by acquiring a linear-in-$q$ dispersion with coefficient $v_s$ in the limit $\lambda_{\rm ph}\rightarrow 0$. 

\begin{figure}
    \centering
    \includegraphics[width=\columnwidth]{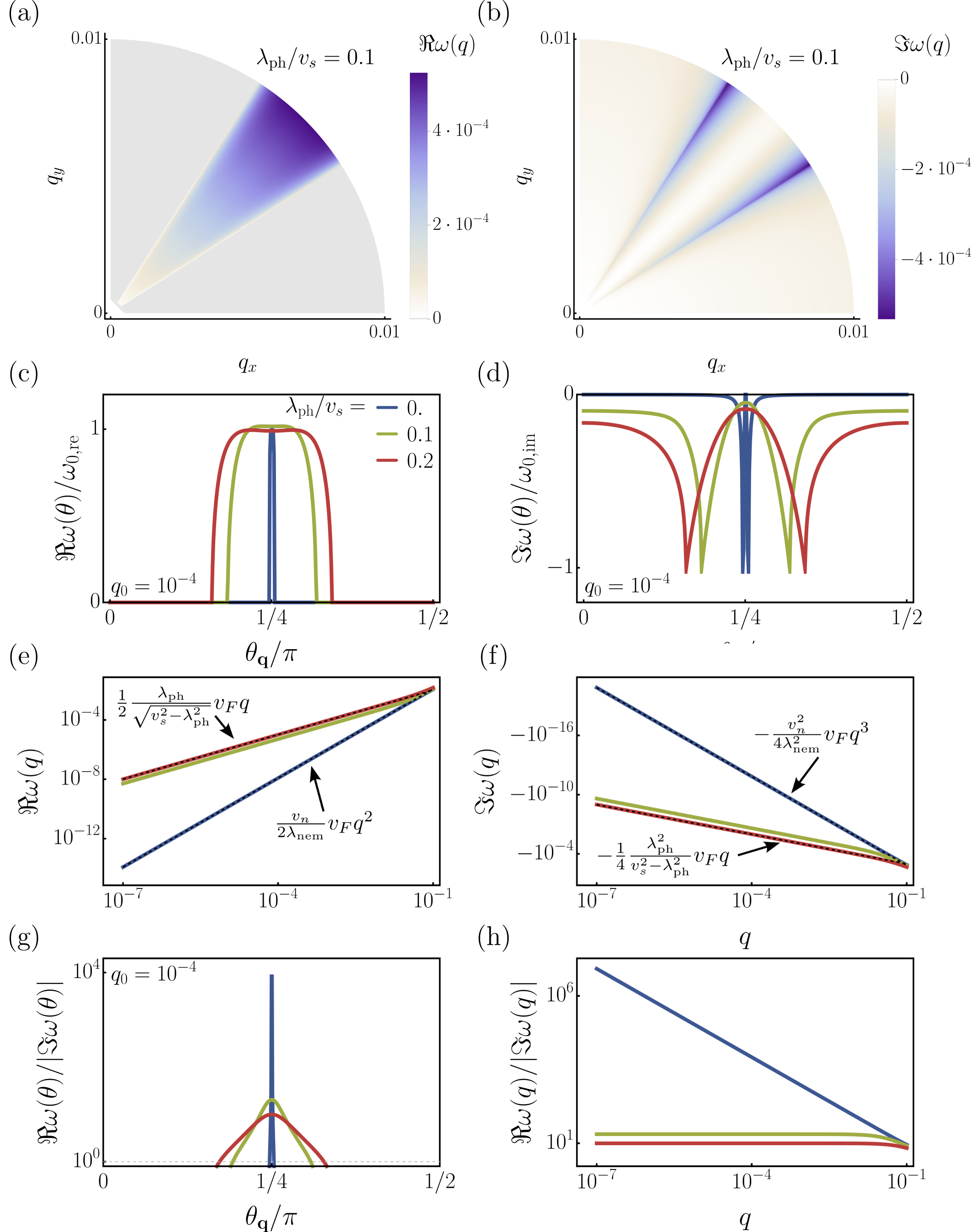}
    \caption{\label{fig:nem_modes_x0}\textbf{Nematic dispersion at the QCP, $x=0$.} Real (a) and imaginary (b) parts of the nematic dispersion for $\lambda_{\rm ph}/v_s =0.1$ plotted as a function of $q_x$ and $q_y$. Here, a gray color denotes zero, showing that the real part of the dispersion is only finite in a region around $\theta_{\mbf{q}}=\tfrac{\pi}{4}$, and the imaginary part exhibits a peak where the real part goes to zero, like in the case where the lattice is uncoupled. (c)--(d) shows the real and imaginary parts of the dispersion as a function of $\theta_{\mbf{q}}$ for fixed $q_0$ as $\lambda_{\rm ph}/v_s$ is varied. Note that the curves are normalized similarly to the ones in Fig.~\ref{fig:nematic_fluctuations_only}(a)--(b). In (e)--(f), the real and imaginary parts of the nematic dispersion are shown as a function of $q$ for fixed $\theta_{\mbf{q}}=\tfrac{\pi}{4}$. Importantly, both real and imaginary parts are linear in $q$ except for the case $\lambda_{\rm ph}/v_s=0$, which corresponds to the uncoupled critical mode [see Figs.\ref{fig:nematic_fluctuations_only}(d)-(e)]. (g)--(h) shows the ratios between the real and the imaginary parts as a function of (g) $\theta_{\mbf{q}}$ and (h) $q$.}
\end{figure}

\begin{figure}
    \centering
    \includegraphics[width=\columnwidth]{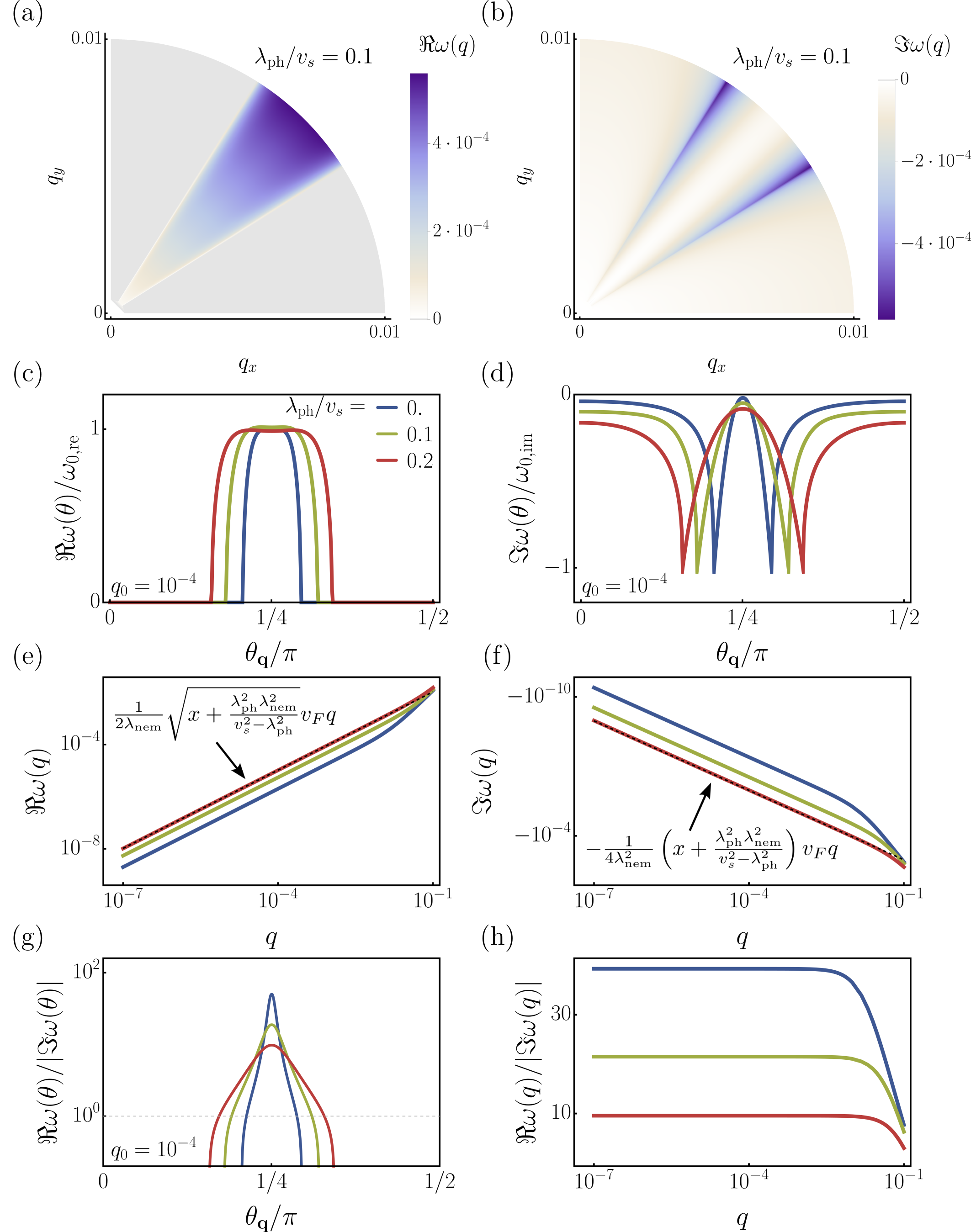}
    \caption{\label{fig:nem_modes_x10-4}\textbf{Nematic dispersion away from the QCP, $x=10^{-4}$.} Real (a) and imaginary (b) parts of the nematic dispersion for $\lambda_{\rm ph}/v_s =0.1$ plotted as a function of $q_x$ and $q_y$. Here, a gray color denotes zero and the figures closely resemble Figs.~\ref{fig:nem_modes_x0}(a)--(b). (c)--(d) shows the real and imaginary parts of the dispersion as a function of $\theta_{\mbf{q}}$ for fixed $q_0$ as $\lambda_{\rm ph}/v_s$ is varied. We normalized the curves like the ones in Fig.~\ref{fig:nematic_fluctuations_only}(a)--(b). In (e)--(f), the real and imaginary parts of the nematic dispersion are shown as functions of $q$ for fixed $\theta_{\mbf{q}}=\tfrac{\pi}{4}$. (g)--(h) shows the ratios between the real and the imaginary parts as a function of (g) $\theta_{\mbf{q}}$ and (h) $q$.}
\end{figure}

\begin{figure}
    \centering
    \includegraphics[width=\columnwidth]{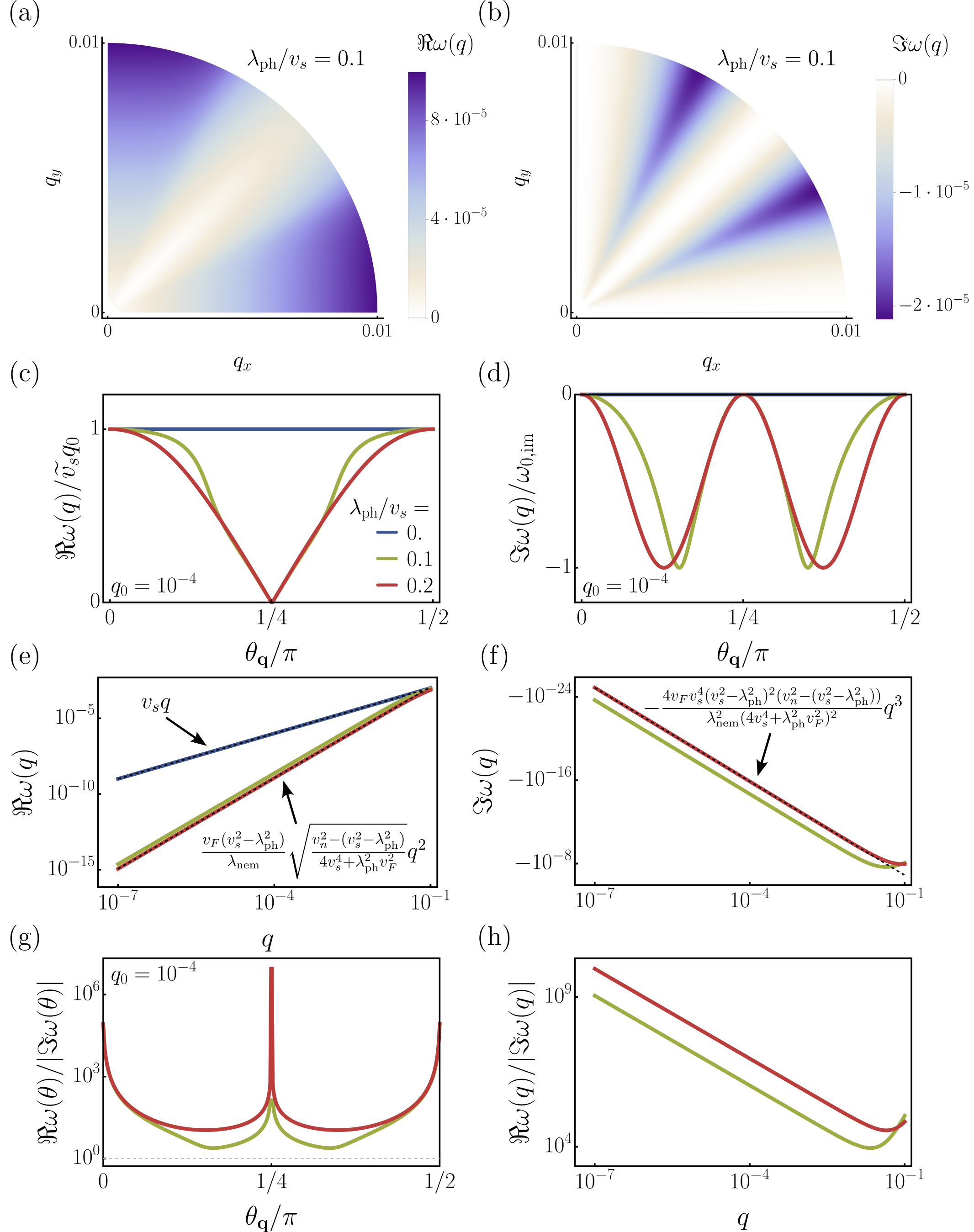}
    \caption{\label{fig:ph_modes_x0}\textbf{ Phonon dispersion at the QCP, $x=0$.} Real (a) and imaginary (b) parts of the phonon dispersion for $\lambda_{\rm ph}/v_s =0.1$ plotted as a function of $q_x$ and $q_y$. (c)--(d) shows the real and imaginary parts of the dispersion as functions of $\theta_{\mbf{q}}$ for fixed $q_0$ as $\lambda_{\rm ph}/v_s$ is varied. Here, the real part is normalized to the $\theta_{\mbf{q}}=0$ value of the dispersion, $\tilde{v}_s q$, where $\tilde{v}_s=\sqrt{v_s^2 + \lambda_{\rm ph}^2}$, while the imaginary part is normalized such that its minimum happens at $-1$. In (e)--(f), the real and imaginary parts of the nematic dispersion are shown as functions of $q$ for fixed $\theta_{\mbf{q}}=\tfrac{\pi}{4}$. Note that the imaginary part for the $\lambda_{\rm ph}/v_s=0$ case vanishes identically. Crucially, the phonon mode exhibits the critical behavior seen in the nematic mode in the uncoupled case: For finite values of $\lambda_{\rm ph}$, the real part is quadratic in $q$ while the imaginary part is cubic in $q$. (g)--(h) shows the ratios between the real and the imaginary parts as a function of (g) $\theta_{\mbf{q}}$ and (h) $q$.}
\end{figure}

\begin{figure}
    \centering
    \includegraphics[width=\columnwidth]{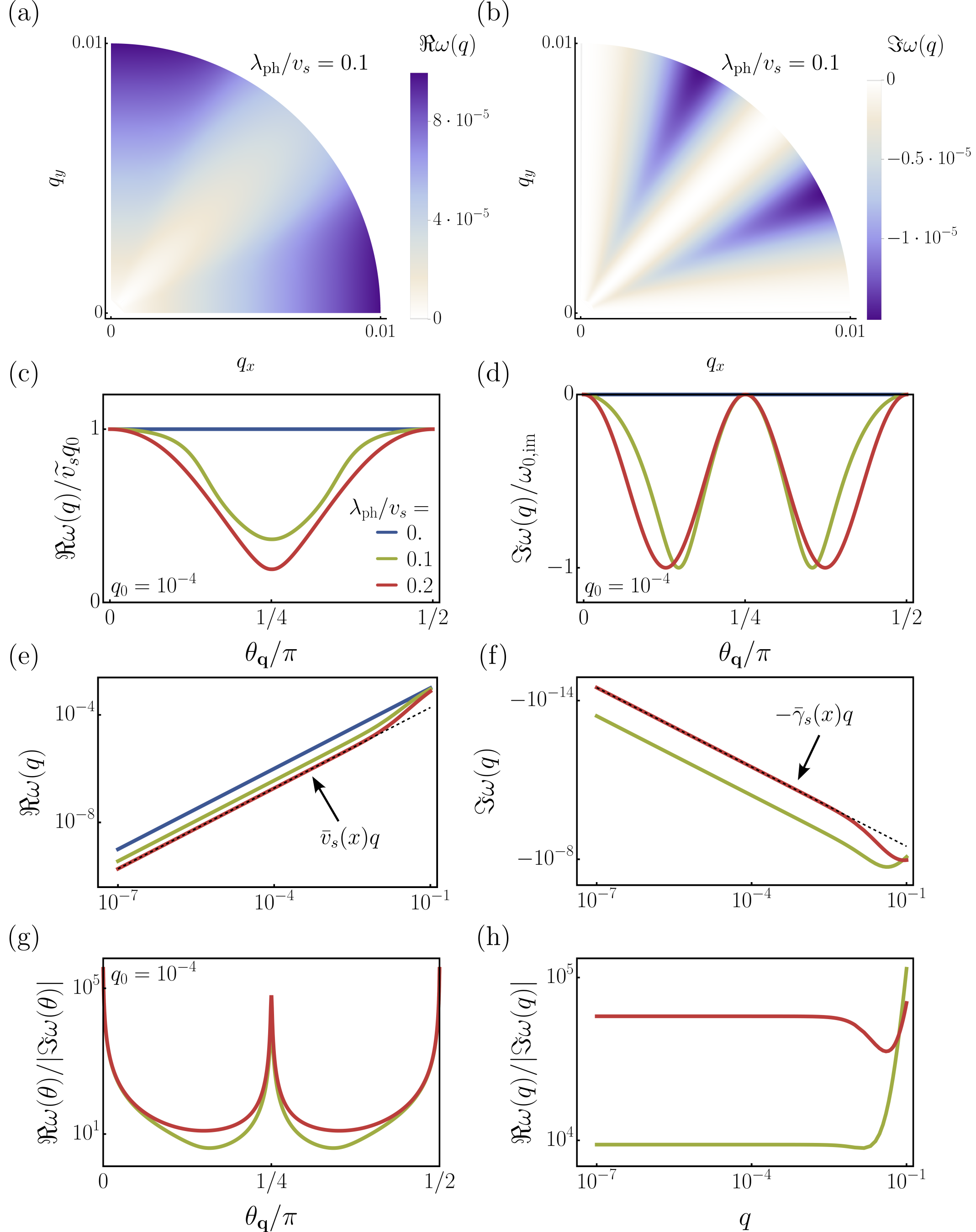}
    \caption{\label{fig:ph_modes_x10-4}\textbf{Phonon dispersion away from the QCP, $x=10^{-4}$.} Real (a) and imaginary (b) parts of the phonon dispersion for $\lambda_{\rm ph}/v_s =0.1$ plotted as a function of $q_x$ and $q_y$. (c)--(d) shows the real and imaginary parts of the dispersion as functions of $\theta_{\mbf{q}}$ for fixed $q_0$ as $\lambda_{\rm ph}/v_s$ is varied. The normalization is identical to the one used in Figs.~\ref{fig:ph_modes_x0}(c)--(d). In (e)--(f), the real and imaginary parts of the nematic dispersion are shown as function sof $q$ for fixed $\theta_{\mbf{q}}=\tfrac{\pi}{4}$. Note that the imaginary part for the $\lambda_{\rm ph}/v_s=0$ case vanishes identically. Both real and imaginary parts are linear in $q$, and the form of the coefficients $\bar{v}_s(x)$ and $\bar{\gamma}_s(x)$ are given in Appendix~\ref{app:analytic_modes}. (g)--(h) shows the ratios of the real and the imaginary parts as a function of (g) $\theta_{\mbf{q}}$ and (h) $q$.}
\end{figure}

The on-shell modes can be found from the poles of the propagators, Eqs.~\eqref{eq:full_ph_prop} and \eqref{eq:full_nem_prop} or, alternatively, from the eigenvalues of Eq.~\eqref{eq:prop_matrix}. In Figs.~\ref{fig:nem_modes_x0}--\ref{fig:ph_modes_x10-4}, we show the nematic and phononic modes resulting from numerically solving for the zeros of the inverse propagators, Eqs.~\eqref{eq:full_ph_prop} and \eqref{eq:full_nem_prop}, for different values of $\lambda_{\rm ph}$ and $x$. The modes shown are the ones with positive real part and negative imaginary part. Here, we focus on the case where the mode-mode coupling mentioned in Sec.~\ref{sec:spectral_functions} can be neglected, such that one can still characterize a mode as either nematic or phononic, based on the above designation (see Appendix~\ref{app:mode_mode} for further details).

Fig.~\ref{fig:nem_modes_x0} illustrates the behavior of the nematic mode at the QCP, corresponding to $x=0$. The combined angular and $q$-dependence of the real and imaginary parts of the dispersion are shown respectively in Fig.~\ref{fig:nem_modes_x0}(a)--(b) for $\lambda_{\rm ph}/v_s=0.1$. They demonstrate that the main qualitative variation in the mode occurs in the angular direction, shown in Fig.~\ref{fig:nem_modes_x0}(c)--(d). As $\lambda_{\rm ph}/v_s$ is increased, both real and imaginary parts increase in magnitude, and the angular region of underdamped behavior becomes wider. Clearly, the effect of a non-zero coupling to the lattice at the QCP is similar to the effect that moving away from the QCP has in the uncoupled limit, as shown in Fig.~\ref{fig:nematic_fluctuations_only}. 
The implication, which we substantiate in the next paragraph, is that the nematic mode, whose damping vanishes at $x=0, q\to 0$ along $\theta_{\mbf{q}}=\pi/4$
in the uncoupled case, acquires a finite damping term (it is still weakly damped but now has a finite lifetime), see Fig. \ref{fig:nem_modes_x0}.

The main difference lies in the $q$-dependence of the mode at the QCP, at $x=0$ , shown in Figs.~\ref{fig:nem_modes_x0}(e)--(f). In the absence of a coupling to the lattice degrees of freedom, the dispersion is qualitatively different at the QCP along the diagonal direction, with the real part being quadratic in $q$ rather than linear and the imaginary part being cubic in $q$ rather than linear [see Eqs.~\eqref{eq:omega_nem_linear_q_no_ph} and \eqref{eq:omega_nem_higher_q_no_ph} and Figs.~\ref{fig:nematic_fluctuations_only}(d)--(e)]. Once the coupling to the lattice is included, the nematic dispersion is modified and, to lowest order in $q$, it is
\begin{align}
    \omega_{\rm nem}(q) &\approx \frac{1}{2\lambda_{\rm nem}}\sqrt{x + \lambda_{\rm p-n}^2} v_F q \nonumber \\ &- i \frac{1}{4\lambda_{\rm nem}^2} \left( x + \lambda_{\rm p-n}^2 \right) v_F q \label{eq:nematic_mode_linear_approximation}
\end{align}
as shown in Appendix~\ref{app:analytic_modes}. The implication of this can be seen in Fig.~\ref{fig:nem_modes_x0}(h) which shows the $q$-dependence of the ratio between the real and imaginary parts of the dispersion along the diagonal. Only the case $\lambda_{\rm ph}=0$ shows the $q^{-1}$-behavior associated with a critical mode in Fig.~\ref{fig:nematic_fluctuations_only}(f). The change of behavior of the dispersion associated with the QCP, as evidenced in Fig.~\ref{fig:nematic_fluctuations_only}, does not emerge in the nematic mode once the coupling to the lattice is included. Instead, this behavior will emerge in the transverse phonon, as demonstrated below. This is consistent with the conclusions presented in Sec.~\ref{sec:adler}: Since the nematic mode is coupled to a Goldstone mode, it will exhibit none of the behavior associated with a critical mode.

The behavior of the nematic mode in the case of a finite $x$ (i.e., away from the QCP) is depicted in Fig.~\ref{fig:nem_modes_x10-4}. The main impact of moving away from the QCP is that, regardless of the value of $\lambda_{\rm ph}$, the curves look qualitatively similar. The focusing effect discussed previously is still apparent here, as seen in Fig.~\ref{fig:nem_modes_x10-4}(g), with smaller values of $\lambda_{\rm ph}/v_s$ leading to a narrower region around the diagonal in which the mode is underdamped. However, as is clear from Eq.~\eqref{eq:nematic_mode_linear_approximation}, the $q$-dependence remains linear in all cases and consequently, the damping tends to a constant as $q \rightarrow 0$, as shown in Fig.~\ref{fig:nem_modes_x10-4}(h).

The critical behavior expected in the system at $x=0$ is exhibited by the phonon-like mode, whose behavior is depicted in Fig.~\ref{fig:ph_modes_x0}. As expected, in the case $\lambda_{\rm ph}=0$, the phonon mode is perfectly coherent (infinitesimal imaginary part) with a dispersion $\omega(q)=v_s q$. The coupling to the electrons, and thereby to the nematic fluctuations, significantly alters this mode, in particular at the QCP. As in the case of the nematic mode, the angular dependence of the phonon mode remains similar for different values of $q$, as evidenced by the combined angular and $q$-dependence in Figs.~\ref{fig:ph_modes_x0}(a)--(b). The angular behavior is highlighted in Figs.~\ref{fig:ph_modes_x0}(c)--(d), and is qualitatively distinct from the nematic case. By assumption, the phonon propagation is isotropic at $\lambda_{\rm ph}=0$. However, for finite values of $\lambda_{\rm ph}/v_s$, the mode becomes anisotropic and, as highlighted in the Introduction, softens first along the diagonal direction. Moreover, the mode is underdamped in the entire region, most significantly along the $\theta_{\mbf{q}}=0, \tfrac{\pi}{2}$ and the $\theta_{\mbf{q}}=\tfrac{\pi}{4}$ directions. At finite $\lambda_{\rm ph}$, while the nematic mode remains linear in $q$ even at $x=0$, the behavior of the phonon mode changes drastically at $x=0$, as shown in Fig.~\ref{fig:ph_modes_x0}(e)--(f). In this case, as shown in Appendix~\ref{app:analytic_modes}, the real part of the mode is quadratic in $q$ while the imaginary part is cubic in $q$:
\begin{align}
    \omega_{\rm ph}(q)|_{x=0} \approx & \frac{v_F(v_s^2 -\lambda_{\rm ph}^2)}{\lambda_{\rm nem}}\sqrt{\frac{v_n^2 - (v_s^2 -\lambda_{\rm ph}^2)}{4v_s^4 + \lambda_{\rm ph}^2v_F^2 }}q^2 \nonumber \\ & - i \frac{4 v_F v_s^4 (v_s^2 - \lambda_{\rm ph}^2)^2(v_n^2 - (v_s^2 -\lambda_{\rm ph}^2))}{\lambda_{\rm nem}^2(4 v_s^4 + \lambda_{\rm ph}^2 v_F^2)^2}q^3
\end{align}
As a consequence, the ratio between the real and the imaginary part diverges as $q \rightarrow 0$ [see Fig.~\ref{fig:ph_modes_x0}(h)], similar to the case of the pure nematic mode studied in Sec.~\ref{sec:nematic_fluctuations}. Hence, this corresponds to the critical mode for all finite values of $\lambda_{\rm ph}/v_s$.

The phonon mode in the case of a finite $x$ is shown in Fig.~\ref{fig:ph_modes_x10-4}. Here, as shown in Figs.~\ref{fig:ph_modes_x10-4}(e)--(f) the phonon mode remains linear in $q$ regardless of the value of $\lambda_{\rm ph}/v_s$, with dispersion
\begin{align}
    \omega_{\rm ph}(q) \approx \bar{v}_s(x) q - i \bar{\gamma}_s (x) q
\end{align}
where $\bar{v}_s(x)$ and $\bar{\gamma}_s (x)$ are functions of $x$ vanishing at $x=0$. Their expressions are given in Appendix~\ref{app:analytic_modes}. While the damping has the same qualitative angular dependence, shown in Fig.~\ref{fig:ph_modes_x10-4}(g), its $q$-dependence is markedly different and tends to a constant as $q \rightarrow 0$ [see Fig.~\ref{fig:ph_modes_x10-4}(h)].

\section{Conclusions}\label{sec:conclusions}

In this work, we presented a derivation of the nemato-elastic coupling from a microscopic model, for which the coupling is mediated by the electronic degrees of freedom. In particular, the nematic fluctuations couple to the lattice vibrations indirectly \emph{via} the coupling between electrons and transverse acoustic phonons. Since the latter vanishes in the simplest electron-phonon process due to the phonon's transverse nature, we invoke the mechanism proposed in Ref. ~\cite{Tsuneto1961Ultrasound,Schmid1973Electron-phonon}, which involves the impact of phonons on the impurity potential, Eq.~\eqref{eq:el_ph_coupling}. This formalism enables us to assess the hybridized nematic-phonon modes as a putative nematic QCP is approached. Due to the coupling to the gapless particle-hole excitations of the metal, an overdamped nematic mode that becomes underdamped at the QCP along certain directions emerges, governing the critical properties of the nematic instability in the absence of the lattice degrees of freedom. These nematic-plasmonic modes hybridize with the massless lattice modes -- the transverse phonons --  giving rise to two hybrid modes. 

These hybrid modes exhibit a mixture of the dynamics characteristic of nematic fluctuations and of transverse phonons. Our systematic investigation of the on-shell dispersions of these hybrid modes as functions of both $q$ and $\theta_{\mbf{q}}$ highlights distinct momentum-space regions of coherent and incoherent modes. Importantly, only one of the two hybrid modes exhibit the characteristic signatures of a critical mode, which is a consequence of Adler's theorem, here manifesting in the massless modes. These results are consistent with the observed lack of divergent effective masses in S-doped FeSe near the nematic critical point~\cite{Reiss2020Quenched,Coldea2021Electronic}.

These findings demonstrate the importance of treating both structural and nematic degrees of freedom on an equal footing to capture their hybridized dynamics. Since unconventional superconductivity near the QCP arises from the exchange of dynamic collective excitations, it will be interesting to employ this formalism to gain further insights into the problem of pairing mediated by quantum critical nematic fluctuations. This is particularly timely given the recent experiments reporting strong evidence for the realization of quantum critical nematic pairing in S-doped FeSe~\cite{SilvaNeto2025}.

\begin{acknowledgments}
We thank A. Chubukov, J. Paaske, and J. Schmalian for fruitful discussions. M.H.C acknowledges support by ERC grant project 101164202 -- SuperSOC. Funded by the European Union. Views and opinions expressed are however those of the authors only and do not necessarily reflect those of the European Union or the European Research Council Executive Agency. Neither the European Union nor the granting authority can be held responsible for them. A.K. acknowledges support by the  United States - Israel Binational Science Foundation (BSF), grant No. 2022242.
\end{acknowledgments}

\appendix

\begin{widetext}
\section{Nematic modes in the absence of a coupling to the lattice}\label{app:nematic_modes_details}

The correction to the self-energy of the nematic fluctuations caused by virtual electron-hole fluctuations is given by the nematic polarization bubble. To one-loop order, this is
\begin{align}
    \Pi_{\rm nem}(\mbf{q},\omega) &= \tilde{\lambda}_{\rm nem}^2 \int \frac{\mathrm{d}^2 k}{(2\pi)^2} k^4 \cos^2 2\theta_{\mbf{k}} \frac{\tanh \left(\tfrac{\beta \epsilon_{\mbf{k}-\mbf{q}/2}}{2}\right) - \tanh \left(\tfrac{\beta \epsilon_{\mbf{k}+\mbf{q}/2}}{2}\right)}{\omega + i 0^+ + \epsilon_{\mbf{k}-\mbf{q}/2} - \epsilon_{\mbf{k}+\mbf{q}/2}} \\
    & \approx \frac{\tilde{\lambda}_{\rm nem}^2 k_{\rm F}^4 d(\epsilon_{\rm F})}{4\pi} \int_0^{2\pi} \mathrm{d}\phi \left(\cos^2 2\theta_{\mbf{q}} -4 \cos 4\theta_{\mbf{q}} \cos^2 \phi \sin^2 \phi \right) \frac{v_{F} q \cos \phi}{\omega + i0^+ -v_{F} q \cos \phi}\,,
\end{align}
where we have expanded for $|\mbf{q}|$ close to $|\mbf{k}|$ with $\phi$ denoting the angle between $\mbf{k}$ and $\mbf{q}$. The term in the parentheses comes from $\theta_{\mbf{k}} = \theta_{\mbf{q}} + \phi$ and ignoring a term that is odd in $\phi$ and therefore integrates to zero. Defining $\zeta = \cos \phi$ and $z=\tfrac{\omega}{v_{F}q}$ we find
\begin{align}
    \Pi_{\rm nem}(\mbf{q},\omega) = \frac{\tilde{\lambda}_{\rm nem}^2 k_{\rm F}^4 d(\epsilon_{\rm F})}{2\pi} \int_{-1}^1 \frac{\mathrm{d}\zeta}{\sqrt{1-\zeta^2}} \left[\cos^2 2\theta_{\mbf{q}} - 4 \cos 4\theta_{\mbf{q}} \zeta^2 \left( 1-\zeta^2 \right) \right] \frac{ \zeta}{z + i 0^+ - \zeta}\,. \label{eq:nematic_polarization_bubble_integral}
\end{align}
The integral depends on whether $|z|<1$ or $|z| >1$, i.e., whether $\omega < v_{F} q$ or $\omega > v_{F} q$. We find
\begin{align}
    \Pi_{\rm nem}(z) = 
    \begin{cases} 
    -\frac{\tilde{\lambda}_{\rm nem}^2 k_{\rm F}^4 d(\epsilon_{\rm F})}{2} \left[ 2 \cos^2 2 \theta_{\mbf{q}} \left( 1 + \frac{i z}{\sqrt{1-z^2}} \right) - \cos 4 \theta_{\mbf{q}} \left(1 + 4z^2 - 8z^4 + i 8 z^3 \sqrt{1-z^2} \right) \right] & z<1 \\
    -\frac{\tilde{\lambda}_{\rm nem}^2 k_{\rm F}^4 d(\epsilon_{\rm F})}{2} \left[ 2 \cos^2 2 \theta_{\mbf{q}} \left( 1 - \frac{z}{\sqrt{z^2-1}} \right) - \cos 4 \theta_{\mbf{q}} \left(1 + 4z^2 - 8z^4 +  8 z^3 \sqrt{z^2-1} \right) \right] & z>1
    \end{cases}\,.\label{eq:nem_polarization_bubble_full}
\end{align}
In the limit of both $\omega$ and $q$ going to zero, the case $\omega < v_{F} q$ leads to modes with $\omega(q \rightarrow 0)=0$ which remain overdamped except at $r = \frac{\tilde{\lambda}_{\rm nem}^2 k_{\rm F}^4 d(\epsilon_{\rm F})}{2}$ (see Fig.~\ref{fig:nematic_fluctuations_only}). As $\omega(q \rightarrow 0)=0$, these modes are formally massless. However, as they are overdamped away from the critical point, there are no coherent excitations as $q \rightarrow 0$, and we prefer the terminology overdamped.

In contrast, the case $\omega > v_{F} q$ allows for massive modes for which $\omega(q \rightarrow 0)=\sqrt{r}$. From Eq.~\eqref{eq:nem_polarization_bubble_full} it is clear that the imaginary part vanishes for $z>1$ and these are long-lived excitations.

In the main text, we focus on the dispersions of the overdamped nematic modes which become underdamped at the critical point. This provides a useful measure of critical behavior of these modes, also in the case where the coupling to the lattice is included, and the massive underdamped mode remains massive even at criticality.

\section{Mode-mode coupling}\label{app:mode_mode}

When the two poles of the propagator appear in close proximity in the complex plane, they exhibit an avoided crossing and the separation into on-shell ``nematic'' and on-shell ``phononic'' modes is no longer clear cut. We illustrate the issue in Fig.~\ref{fig:mode_mode_example} which shows the evolution of the poles of the nematic propagator as a function of $\theta_{\mbf{q}}$ for $\lambda_{\rm ph}/v_s=0.06$ (in light brown) and $\lambda_{\rm ph}/v_s=0.07$ (in dark green) for $x=0$ and $q=q_0=10^{-4}$. The green case corresponds to the usual situation: The nematic mode, corresponding to the large semi-circle starting near $\Re \omega(\theta_{\mbf{q}}) \approx 3.5 \cdot 10^{-6}$, looks very similar to the nematic mode obtained in the $\lambda_{\rm ph}=0$ case. The phonon mode starts near $(0,0)$ (on this scale) and terminates at $v_s q_0= 1 \cdot 10^{-6}$. The light brown case is qualitatively different. Here, the poles exhibit an avoided crossing near the imaginary axis and one of the poles leaves the imaginary axis again and terminates at $v_s q_0$, despite having started out deep in the complex plane, in this case near $\Re \omega(\theta_{\mbf{q}}) \approx 3 \cdot 10^{-6}$. Due to this avoided crossing, applying the definition from Sec.~\ref{sec:on-shell} to separate the two modes will lead to jumps in the dispersions. Consequently, in Figs.~\ref{fig:nem_modes_x0}--\ref{fig:ph_modes_x10-4}, we consider only parameters where this is not an issue.

\begin{figure}
    \centering
    \includegraphics[width=0.5\columnwidth]{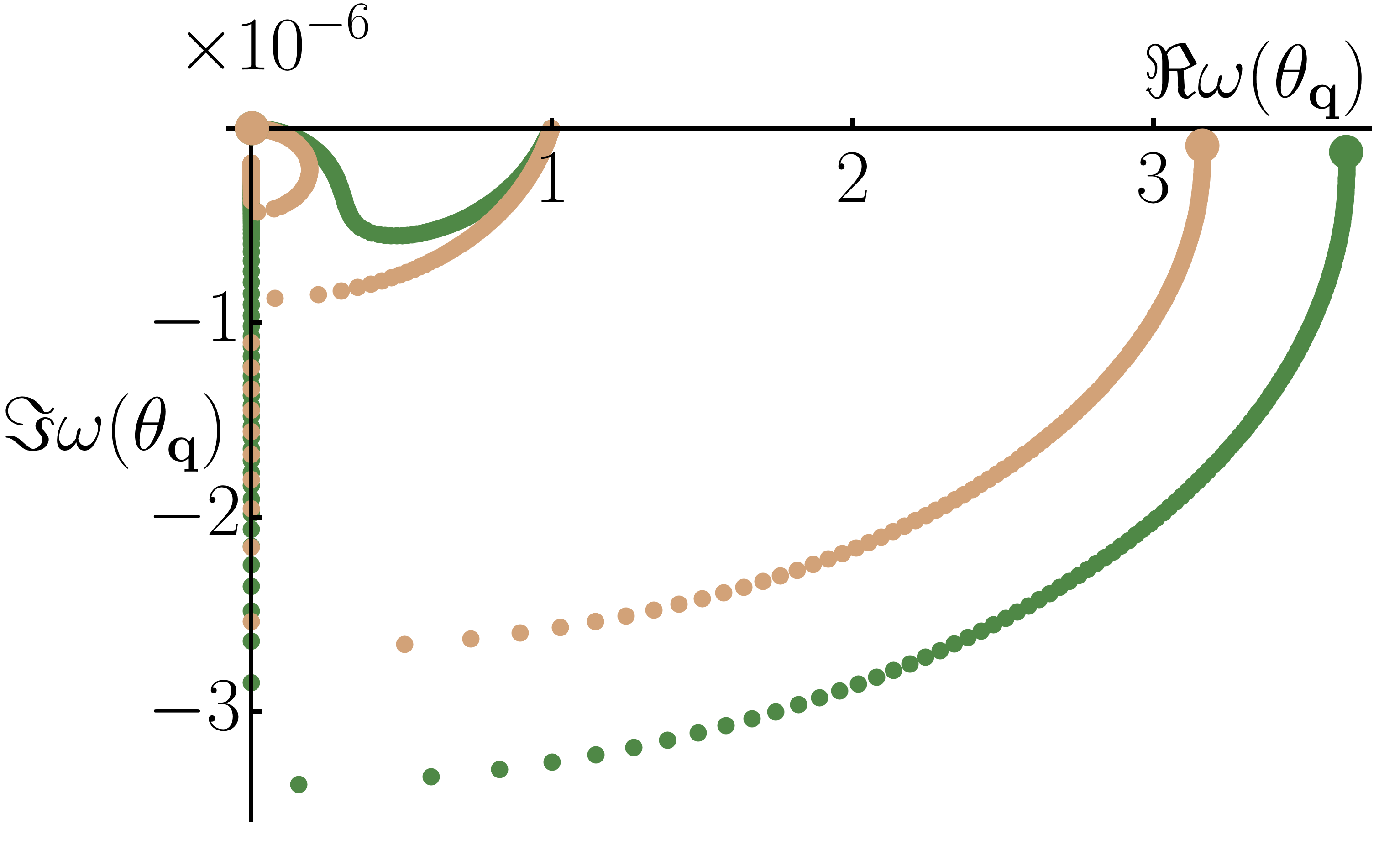}
    \caption{\label{fig:mode_mode_example} \textbf{Illustration of the mode-mode coupling.} Movement of the poles of the full ``nematic'' propagator as a function of $\theta_{\mbf{q}}$ in the complex plane with larger dots denoting the $\theta_{\mbf{q}}=\tfrac{\pi}{4}$ pole. Here, $x=0$ and $q=q_0=10^{-4}$. The green dots correspond to the $\lambda_{\rm ph}/v_s=0.07$ case where the poles are sufficiently far apart that the definition of Sec.~\ref{sec:on-shell} can be applied. The light brown dots correspond to $\lambda_{\rm ph}/v_s=0.06$ in which case the behaviour of the poles change qualitatively.}
\end{figure}

\section{Analytical approximation for the on-shell modes}\label{app:analytic_modes}

To gain some analytical insight into the impact of $\lambda_{\rm ph}$ on the collective hybridized nematic-phonon modes, it is convenient to use the eigenvalues of Eq.~\eqref{eq:prop_matrix} (in the absence of the $-\omega^2 \mathds{1}$ part which will be reinstated later) to determine the poles. These modes are
\begin{equation}
    E_{\pm} = \frac{1}{2}\left( \widetilde{\chi}^{-1}_{\rm dres}(\mbf{q},\omega) + \widetilde{\mathcal{D}}_{\rm dres}^{-1}(\mbf{q},\omega)  \pm \left|\widetilde{\chi}^{-1}_{\rm dres}(\mbf{q},\omega) - \widetilde{\mathcal{D}}_{\rm dres}^{-1}(\mbf{q},\omega)\right|\sqrt{ 1 - \frac{4\Pi_{\rm ph-nem}(\mbf{q},\omega)^2}{\left[ \widetilde{\chi}^{-1}_{\rm dres}(\mbf{q},\omega) - \widetilde{\mathcal{D}}_{\rm dres}^{-1}(\mbf{q},\omega) \right]^2}} \right)\,,\label{eq:eigenvalues}
\end{equation}
where, for convenience, $\widetilde{\chi}^{-1}_{\rm dres}(\mbf{q},\omega)$ and $\widetilde{\mathcal{D}}^{-1}_{\rm dres}(\mbf{q},\omega)$ refer to the dressed propagators without the $-\omega^2$-term:
\begin{align}
    \widetilde{\chi}^{-1}_{\rm dres}(\mbf{q},\omega) &= r+v_n^2 q^2 + \Pi_{\rm nem}(\mbf{q},\omega)\\
    \widetilde{\mathcal{D}}^{-1}_{\rm dres}(\mbf{q},\omega) &= v_s^2 q^2 + \Pi_{\rm ph}(\mbf{q},\omega)
\end{align}
Expanding Eq.~\eqref{eq:eigenvalues} in powers of $\lambda_{\rm ph}^2\lambda_{\rm nem}^2$ to lowest order, we obtain
\begin{align}
    E_{\rm nem} &\approx \widetilde{\chi}_{\rm dres}^{-1}(\mbf{q},\omega) - \frac{\Pi_{\rm ph-nem}(\mbf{q},\omega)^2}{\widetilde{\chi}^{-1}_{\rm dres}(\mbf{q},\omega)-\widetilde{\mathcal{D}}^{-1}_{\rm dres}(\mbf{q},\omega)}\,, \label{eq:E_nem_approx} \\
    E_{\rm ph} &\approx \widetilde{\mathcal{D}}_{\rm dres}^{-1}(\mbf{q},\omega) + \frac{\Pi_{\rm ph-nem}(\mbf{q},\omega)^2}{\widetilde{\chi}^{-1}_{\rm dres}(\mbf{q},\omega)-\widetilde{\mathcal{D}}_{\rm dres}^{-1}(\mbf{q},\omega)}\,, \label{eq:E_ph_approx}
\end{align}
where, without loss of generality, we changed from $E_{\pm}$ to $E_{\rm nem}$ and $E_{\rm ph}$ to emphasize their relation to the unperturbed propagators. Here, $E_{\rm nem}$ originates from $E_+$ if $\widetilde{\chi}_{\rm dres}^{-1}(\mbf{q},\omega) > \widetilde{\mathcal{D}}_{\rm dres}^{-1}(\mbf{q},\omega)$ and from $E_{-}$ if $\widetilde{\chi}_{\rm dres}^{-1}(\mbf{q},\omega) < \widetilde{\mathcal{D}}_{\rm dres}^{-1}(\mbf{q},\omega)$ and vice versa for $E_{\rm ph}$. Focusing on the direction $\theta_{\mbf{q}}=\tfrac{\pi}{4}$, we let $r \rightarrow x + \lambda_{\rm nem}^2 \left(1 + \frac{\lambda_{\rm ph}^2}{v_s^2 - \lambda_{\rm ph}^2} \right)$  and expand Eqs.~\eqref{eq:E_nem_approx} and \eqref{eq:E_ph_approx} to third order in $\tfrac{\omega}{v_F q}$ and reinstate the diagonal $-\omega^2$-term resulting in two equations for the dispersion of the nematic and phononic mode, respectively,
\begin{align}
    x + \frac{\lambda_{\rm ph}^2 \lambda_{\rm nem}^2}{v_s^2 - \lambda_{\rm ph}^2} + [v_n^2 + \alpha(q)]q^2 - \left(1 + \frac{4\lambda_{\rm nem}^2 [1+\lambda_{\rm ph}^2 \kappa(q) q^2]}{v_F^2 q^2} \right)\omega^2 - i\frac{8\lambda_{\rm nem}^2[1+\lambda_{\rm ph}^2 \kappa(q) q^2]}{v_F^3 q^3}\omega^3 = 0 \label{eq:nematic_dispersion_equation} \\
    [v_s^2 -\lambda_{\rm ph}^2 -\alpha(q)]q^2 - \left(1 - \frac{4\lambda_{\rm ph}^2[\kappa(q) \lambda_{\rm nem}^2 - 1]}{v_F^2} \right)\omega^2 + i\frac{8\lambda_{\rm ph}^2[\lambda_{\rm nem}^2 \kappa(q) -1]}{v_F^3 q}\omega^3 = 0\,. \label{eq:phonon_dispersion_equation}
\end{align}
Here, we have defined
\begin{align}
    \beta(q) &= \frac{1}{x + v_n^2 q^2 - (v_s^2 - \lambda_{\rm ph}^2)q^2+ \frac{\lambda_{\rm ph}^2 \lambda_{\rm nem}^2}{v_s^2 -\lambda_{\rm ph}^2}} \\
    \alpha(q) &= \lambda_{\rm ph}^2 \lambda_{\rm nem}^2 \beta(q)\,, \\
    \kappa(q) &= (\lambda_{\rm ph}^2 q^2 - \lambda_{\rm nem}^2)\beta(q)^2 - 2 \beta(q)\,.
\end{align}
Crucially, the $\omega$-independent terms can be treated as perturbations in the long-wavelength limit. In Eq.~\eqref{eq:nematic_dispersion_equation}, the $\omega$-independent terms tend to a constant as $q \rightarrow 0$, while the coefficients of $\omega^2$ and $\omega^3$ both diverge as $q \rightarrow 0$. In Eq.~\eqref{eq:phonon_dispersion_equation}, the $\omega$-independent coefficient goes to zero as $q \rightarrow 0$, while the coefficient of $\omega^2$ tends to a constant and the coefficient of $\omega^3$ diverges.

The solutions, to first order in the $\omega$-independent terms, that have a positive real and a negative imaginary part are
\begin{align}
    \omega_{\rm nem} (q) & \approx \sqrt{\frac{x + \frac{\lambda_{\rm ph}^2 \lambda_{\rm nem}^2}{v_s^2 -\lambda_{\rm ph}^2} + [v_n^2 + \alpha(q)]q^2}{4\lambda_{\rm nem}^2 +[v_F^2 + 4 \lambda_{\rm ph}^2 \lambda_{\rm nem}^2 \kappa(q)]q^2}}v_F q - i \frac{4 \lambda_{\rm nem}^2 [1 + \lambda_{\rm ph}^2 \kappa(q) q^2]\left\{ x + \frac{\lambda_{\rm ph}^2 \lambda_{\rm nem}^2}{v_s^2 -\lambda_{\rm ph}^2} + [v_n^2 + \alpha(q)]q^2\right\}}{[4\lambda_{\rm nem}^2 +[v_F^2 + 4 \lambda_{\rm ph}^2 \lambda_{\rm nem}^2 \kappa(q)]q^2]^2}v_F q \label{eq:omega_nem_w_lambda_ph} \\
    \omega_{\rm ph}(q) & \approx \sqrt{\frac{v_s^2 - \lambda_{\rm ph}^2 - \alpha(q)}{v_F^2 - 4\lambda_{\rm ph}^2[\lambda_{\rm nem}^2 \kappa(q) -1]}} v_F q - i \frac{4\lambda_{\rm ph}^2[v_s^2 - \lambda_{\rm ph}^2 - \alpha(q)][1- \lambda_{\rm nem}^2 \kappa(q)]}{\left\{v_F^2 - 4 \lambda_{\rm ph}^2[\lambda_{\rm nem}^2 \kappa(q) -1] \right\}^2}v_F  q\,. \label{eq:omega_ph_w_lambda_ph}
\end{align}
In the limit where $\lambda_{\rm ph}\rightarrow 0$, the expression in Eq.~\eqref{eq:omega_nem_w_lambda_ph} for the nematic mode reduces to the one of Eq.~\eqref{eq:omega_sol_3}. The simple $q$-dependence in Eqs.~\eqref{eq:omega_nem_w_lambda_ph} and \eqref{eq:omega_ph_w_lambda_ph} is spoiled by the fact that $\alpha$ and $\kappa$ depend on $q$. Away from the QCP we find, to linear order in $q$:
\begin{align}
    \omega_{\rm nem}(q) &\approx \left[ \frac{1}{2\lambda_{\rm nem}}\sqrt{x + \frac{\lambda_{\rm ph}^2 \lambda_{\rm nem}^2}{v_s^2 - \lambda_{\rm ph}^2}} - i \frac{1}{4\lambda_{\rm nem}^2}\left( x+ \frac{\lambda_{\rm ph}^2 \lambda_{\rm nem}^2}{v_s^2 - \lambda_{\rm ph}^2} \right) \right]v_F q \\  
    \omega_{\rm ph}(q) &\approx \sqrt{\frac{x(v_s^2 - \lambda_{\rm ph}^2)\left(x + \frac{\lambda_{\rm ph}^2 \lambda_{\rm nem}^2}{v_s^2 - \lambda_{\rm ph}^2} \right)}{v_F^2  \left(x + \frac{\lambda_{\rm ph}^2 \lambda_{\rm nem}^2}{v_s^2 -\lambda_{\rm ph}^2}\right)^2 + 4\lambda_{\rm ph}^2\left(x + \frac{\lambda_{\rm nem}^2 v_s^2}{v_s^2 - \lambda_{\rm ph}^2} \right)^2}}v_F q \nonumber \\
    & - i \frac{\lambda_{\rm ph}^2 x \left(x + \frac{\lambda_{\rm ph}^2 \lambda_{\rm nem}^2}{v_s^2 - \lambda_{\rm ph}^2} \right) \left(x + \frac{\lambda_{\rm nem}^2 v_s^2}{v_s^2 -\lambda_{\rm ph}^2} \right)^2}{\left( v_F^2 \left(x + \frac{\lambda_{\rm ph}^2\lambda_{\rm nem}^2}{v_s^2 - \lambda_{\rm ph}^2} \right)^2+ 4 \lambda_{\rm ph}^2 \left(x + \frac{\lambda_{\rm nem}^2 v_s^2}{v_s^2 -\lambda_{\rm ph}^2} \right)^2 \right)^2} (v_s^2 - \lambda_{\rm ph}^2)v_F q\,,\label{eq:omega_ph_w_lambda_ph_linear_q}
\end{align}
implying that the nematic mode clearly reduces to Eq.~\eqref{eq:omega_nem_linear_q_no_ph} for $\lambda_{\rm ph}=0$. At the QCP, the linear terms in the phonon dispersion vanish, as expected for a transverse mode at a structural transition. Instead, the leading order term in the real part is quadratic in $q$ while the leading order in the imaginary part is cubic in $q$:
\begin{equation}
    \omega_{\rm ph}(q)|_{x=0} \approx \frac{v_F(v_s^2 -\lambda_{\rm ph}^2)}{\lambda_{\rm nem}}\sqrt{\frac{v_n^2 - (v_s^2 -\lambda_{\rm ph}^2)}{4v_s^4 + \lambda_{\rm ph}^2v_F^2 }}q^2 - i \frac{4 v_F v_s^4 (v_s^2 - \lambda_{\rm ph}^2)^2(v_n^2 - (v_s^2 -\lambda_{\rm ph}^2))}{\lambda_{\rm nem}^2(4 v_s^4 + \lambda_{\rm ph}^2 v_F^2)^2}q^3\,.
\end{equation}
Note that, for $\lambda_{\rm ph}=0$, the phonon mode is unaffected by the QCP, as it is due to critical nematic fluctuations. In this case, the phonon dispersion is purely real and given by $v_s q$ [see Eq.~\eqref{eq:omega_ph_w_lambda_ph_linear_q}], as it should be.

\end{widetext}

\phantom{\cite{christensen_notebook}}

\bibliography{nematic_phonon_bib}

\end{document}